\documentclass[aip,jmp,12pt]{revtex4-1}
\usepackage{amsmath}
\usepackage{amsthm}
\usepackage{amsfonts}
\usepackage{tikz}
\usetikzlibrary{arrows}
\usepackage[colorlinks,citecolor=blue,urlcolor=blue]{hyperref}

\theoremstyle{plain}
\newtheorem{thm}{Theorem}[section]
\newtheorem{lem}[thm]{Lemma}
\newtheorem{prop}[thm]{Proposition}
\newtheorem{cor}[thm]{Corollary}
\newtheorem{conj}[thm]{Conjecture}
\theoremstyle{remark}
\newtheorem{rem}[thm]{Remark}
\theoremstyle{definition}
\newtheorem{defn}{Definition}

\begin{document}

\title{Gaussian optimizers for entropic inequalities in quantum information}

\author{Giacomo De Palma}
\affiliation{QMATH, Department of Mathematical Sciences, University of Copenhagen, Universitetsparken 5, 2100 Copenhagen, Denmark}
\author{Dario Trevisan}
\affiliation{Universit\`a degli Studi di Pisa, 56126 Pisa, Italy}
\author{Vittorio Giovannetti}
\affiliation{NEST, Scuola Normale Superiore and Istituto Nanoscienze-CNR, 56126 Pisa, Italy}
\author{Luigi Ambrosio}
\affiliation{Scuola Normale Superiore, 56126 Pisa, Italy}

\begin{abstract}
We survey the state of the art for the proof of the quantum Gaussian optimizer conjectures of quantum information theory.
These fundamental conjectures state that quantum Gaussian input states are the solution to several optimization problems involving quantum Gaussian channels.
These problems are the quantum counterpart of three fundamental results of functional analysis and probability: the Entropy Power Inequality, the sharp Young’s inequality for convolutions, and the theorem ``Gaussian kernels have only Gaussian maximizers.''
Quantum Gaussian channels play a key role in quantum communication theory: they are the quantum counterpart of Gaussian integral kernels and provide the mathematical model for the propagation of electromagnetic waves in the quantum regime.
The quantum Gaussian optimizer conjectures are needed to determine the maximum communication rates over optical fibers and free space.
The restriction of the quantum-limited Gaussian attenuator to input states diagonal in the Fock basis coincides with the thinning, which is the analog of the rescaling for positive integer random variables.
Quantum Gaussian channels provide then a bridge between functional analysis and discrete probability.
\end{abstract}

\maketitle


\tableofcontents

\section{Introduction}\label{sec:intro}
Gaussian functions play a key role in both functional analysis and probability, and are the solution to several optimization problems involving Gaussian kernels.
The most prominent among these problems is determining the norms of the Gaussian integral kernels $G$ that send a function $f\in L^p(\mathbb{R}^m)$ to the function $Gf\in L^q(\mathbb{R}^n)$ with $p,\,q\ge1$.
In the seminal paper ``Gaussian kernels have only Gaussian maximizers'' \cite{lieb1990Gaussian}, Lieb proved that these norms are achieved by Gaussian functions.
A closely related fundamental result is the sharp Young's inequality for convolutions \cite{beckner1975inequalities,brascamp2002best,fournier1977sharpness,christ_near-extremizers_2011,bennett_brascamp-lieb_2008}, stating that for any $p,\,q,\,r\ge1$, the ratio $\left.\|f*g\|_r\right/\|f\|_p\|g\|_q$ with $f\in L^p(\mathbb{R}^n)$ and $g\in L^q(\mathbb{R}^n)$ is maximized by Gaussian functions, where $f*g$ denotes the convolution of $f$ with $g$.
This inequality has several fundamental applications, such as a proof of the Entropy Power Inequality \cite{lieb1978proof,dembo1991information,cover2006elements}, of the Brunn-Minkowski inequality \cite{gardner2002brunn,cover2006elements} and Lieb's solution \cite{lieb1978proof,lieb2014proof} of Wehrl's conjecture \cite{wehrl1979relation,anderson1993information}, stating that coherent states minimize the Wehrl entropy.
The theorem ``Gaussian kernels have only Gaussian maximizers'' and the sharp Young's inequality for convolutions are among the most important inequalities of functional analysis (see e.g. the book Ref. \onlinecite{lieb2001analysis}).

The Entropy Power Inequality \cite{dembo1991information,stam1959some,shannon2001mathematical} states that the Shannon differential entropy of the sum of two independent random variables with values in $\mathbb{R}^n$ and given Shannon differential entropies is minimum when the two random variables are Gaussian, and is a fundamental tool of information theory \cite{cover2006elements}.
The Entropy Power Inequality was introduced by Shannon to provide an upper bound to the information capacity of non-Gaussian channels \cite{shannon2001mathematical}, and was later used to bound the information capacity region of the Gaussian broadcast channel \cite{bergmans1974simple} and the secret information capacity of the Gaussian wiretap channel \cite{leung1978gaussian}.
The Entropy Power Inequality was also employed to prove the convergence in relative entropy for the central limit theorem \cite{barron1986entropy}.

Quantum information theory \cite{nielsen2010quantum,hayashi2016quantum,wilde2017quantum,holevo2013quantum} is the theory of the transmission and the processing of the information stored in quantum systems.
Most of nowadays communications are made with electromagnetic signals traveling through optical fibers or free space.
Quantum Gaussian channels \cite{caves1994quantum,braunstein2005quantum,weedbrook2012gaussian,holevo2015gaussian} are the quantum counterpart of Gaussian integral kernels, and an $n$-mode quantum Gaussian channel provides the mathematical model for the propagation of $n$ modes of the electromagnetic radiation along an optical fibre or free space in the quantum regime.
For this reason, quantum Gaussian channels play a key role in quantum communication theory.

The subject of this review is the generalization of all the above inequalities for the convolution and for Gaussian integral kernels to quantum Gaussian channels.
The solutions to the resulting quantum optimization problems are conjectured to be quantum Gaussian states, the quantum counterpart of Gaussian probability measures.
This Gaussian optimizer problem arose in quantum information to determine the classical information capacity of phase-covariant quantum Gaussian channels \cite{holevo2001evaluating,giovannetti2004minimum,giovannetti2004minimumb,de2014classical}.
Indeed, proving that the coherent states constitute an optimal coding requires to prove a minimum output entropy conjecture, stating that the coherent input states minimize the output entropy of $n$-mode phase-covariant quantum Gaussian channels (Theorem \ref{thm:MOE}).
This conjecture implies that both the minimum output entropy and the classical capacity of phase-covariant quantum Gaussian channels are additive with respect to the tensor product, i.e., that entanglement does not increase the communication rate.
Moreover, the conjecture also implies the optimality of Gaussian discord \cite{pirandola2014optimality}.
While the minimum output entropy of any classical channel is trivially additive, this property does not hold in general for quantum channels \cite{hastings2009superadditivity}.
The proof of the minimum output entropy conjecture has then been a fundamental result, which required more than ten years \cite{giovannetti2010generalized,garcia2012majorization,garcia2014holy,giovannetti2015solution,mari2014quantum,giovannetti2014ultimate} (see the review Ref. \onlinecite{holevo2015gaussian}; see also Ref. \onlinecite{holevo2016constrained} for the capacity of non phase-covariant quantum Gaussian channels).

Proving that the coherent states constitute an optimal coding for the Gaussian broadcast channel requires a constrained version of the minimum output entropy conjecture.
This constrained version states that quantum Gaussian input states minimize the output entropy of $n$-mode quantum Gaussian channels among all the input states with a given entropy \cite{guha2007classical,guha2007classicalproc,guha2008multiple,guha2008capacity,qi2016capacities} (Conjecture \ref{conj:MOE}).
The constrained minimum output entropy conjecture also implies the converse theorems for the triple trade-off coding with the quantum-limited attenuator and amplifier \cite{wilde2012information,wilde2012quantum,qi2016capacities}.
The conjecture has been generalized to the Entropy Photon-number Inequality \cite{guha2008entropy,guha2008multiple}, stating that quantum Gaussian input states minimize the output entropy of the beam-splitter among all the couple of input states each with a given entropy (Conjecture \ref{conj:EPnI}).
Moreover, it has been realized \cite{de2016gaussiannew} that the constrained minimum output entropy conjecture would follow from the generalization of the theorem ``Gaussian kernels have Gaussian maximizers'' to $n$-mode quantum Gaussian channels (Conjecture \ref{conj:pq}).
Since the beam-splitter is the quantum counterpart of the convolution, the Entropy Photon-number Inequality is the quantum counterpart of the Entropy Power Inequality.
Based on this relation, we conjecture for the first time in this review the validity of a sharp Young's inequality for the beam-splitter (Conjecture \ref{conj:young}).

\begin{figure}[!ht]
\begin{center}
\begin{tikzpicture}[->,>=stealth',shorten >=1pt,auto, thick,
			main node/.style={rounded corners,
			draw=black, thick,
			minimum height=2em,
			inner sep=2pt,
			text centered}]

	\node[main node,fill=red] (young)  {
		\begin{tabular}{c}
			\textbf{Sharp Young's inequality}\\
            \textbf{for the beam-splitter}\\
			{(Conjecture \ref{conj:young}) }
		\end{tabular}};

	\node[main node, fill=yellow] (gausslieb) [right of =young, node distance=20em, yshift=-0.5em, xshift=-1em]  {
		\begin{tabular}{c}
			\textbf{Quantum Gaussian channels}\\
			\textbf{have Gaussian maximizers}\\
			{(Conjecture \ref{conj:pq}) }
		\end{tabular}};

	\node[main node, fill=red] (epni) [below of =young, node distance=8em]  {
		\begin{tabular}{c}
			\textbf{Entropy Photon-number Inequality}\\
			{(Conjecture \ref{conj:EPnI}) }
		\end{tabular}};

 	\node[main node, fill=yellow] (cminent) [below of =gausslieb, node distance=18em, xshift=-5.5em, yshift=-6.5em]  {
		\begin{tabular}{c}
			\textbf{Constrained minumum output entropy}\\
			\textbf{of quantum Gaussian channels}\\
			{(Conjecture \ref{conj:MOE}) }
		\end{tabular}};

 	\node[main node, fill=green] (qepi) [below of =epni, node distance=12em, xshift=-3em, yshift=-4em]  {
		\begin{tabular}{c}
			\textbf{Quantum Entropy}\\
			\textbf{Power Inequality}\\
			{(Theorem \ref{thm:qEPI}) }
		\end{tabular}};

  	\node[main node, fill=green] (minent) [below of =qepi, node distance=6em, xshift=1em, yshift=-4em]  {
		\begin{tabular}{c}
			\textbf{Minimum output entropy of}\\
            \textbf{quantum Gaussian channels}\\
			{(Theorem \ref{thm:MOE}) }
		\end{tabular}};

 	\node[main node, fill=green] (majorization) [right of=minent, node distance=20em, yshift=-3em, xshift=8em]  {
		\begin{tabular}{c}
			\textbf{Majorization for quantum}\\
			\textbf{Gaussian channels}\\
			{(Theorem \ref{thm:maj}) }
		\end{tabular}};
		
 	\node[main node, fill=green] (normpp) [below right of =gausslieb, node distance=11em, xshift=-2em, yshift=-1em]  {
		\begin{tabular}{c}
			\textbf{$p\to p$ norms of}\\
            \textbf{quantum Gaussian channels}\\
			{(Theorem \ref{thm:infty}) }
		\end{tabular}};
		
	\node[main node, fill=green] (norm1p) [below of =normpp, node distance=6em, xshift=-5em, yshift=-2em]  {
		\begin{tabular}{c}
			\textbf{$1\to p$ norms of}\\
            \textbf{quantum Gaussian channels}\\
			{(Corollary \ref{cor:1p}) }
		\end{tabular}};

    \path[every node/.style={font=\small}]
        (young) edge [bend left] (gausslieb)
		(gausslieb) edge  [bend left]  (normpp)
		(gausslieb) edge  [bend right]  (norm1p)
		(majorization) edge  [bend right] (norm1p)
		(gausslieb) edge [bend right]   (cminent)
		(cminent) edge [bend left]   (minent)
		(majorization) edge  [bend left] (minent)
		(young) edge [bend right] (epni)
		(epni) edge [bend right] (cminent)
		(epni) edge [bend right] (qepi);

\end{tikzpicture}
\caption{Implications among conjectures and results.
Green = proven result; Yellow = result proven in some particular cases; Red = open conjecture.}
\label{fig:implications}
\end{center}
\end{figure}
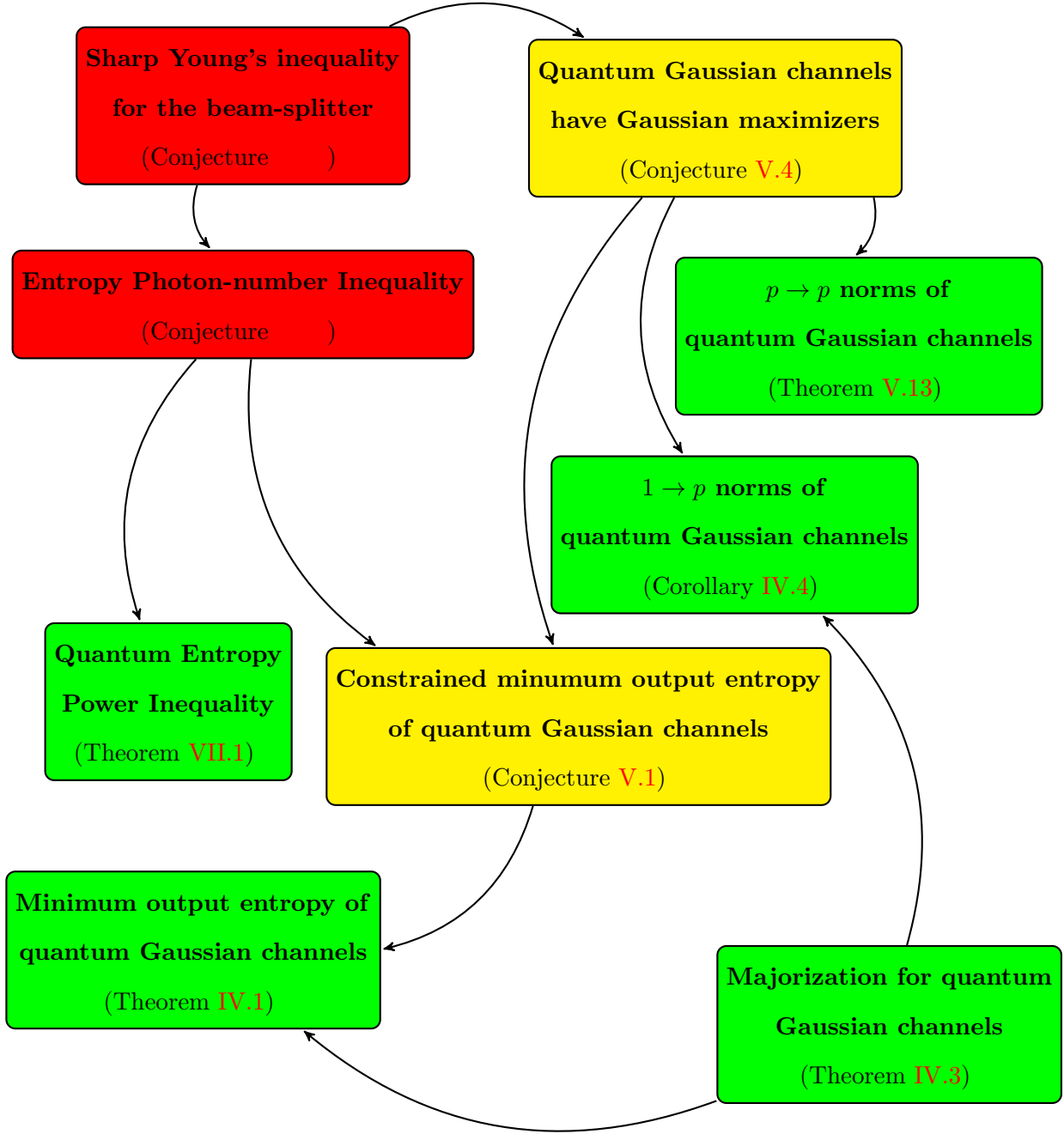

The proof of all the above quantum inequalities has been completed only in some particular cases, and is currently an active field of research.
The constrained minimum output entropy conjecture has been proven only for one-mode quantum Gaussian channels \cite{de2015passive,de2016gaussian,de2017gaussian,de2016gaussiannew,qi2017minimum} or for input states diagonal in some joint product basis \cite{de2017multimode}.
These results are based on a new majorization theorem for one-mode quantum Gaussian channels \cite{de2015passive} (Theorem \ref{thm:maj2}).
The majorization result has been extended to single-jump lossy quantum channels \cite{de2016passive}, but unfortunately it fails for multi-mode quantum Gaussian channels \cite{de2016passive}.
The proof of the constrained minimum output entropy conjecture for one-mode quantum Gaussian channels made possible the proof of the fundamental relation between the von Neumann and the Wehrl entropy, stating that for any $n$, $n$-mode quantum Gaussian states have the minimum Wehrl entropy among all the $n$-mode quantum states with a given von Neumann entropy \cite{de2017wehrl}.
For generic $p,\,q\ge1$, the theorem ``Gaussian kernels have Gaussian maximizers'' has been proven only for one-mode quantum Gaussian channels \cite{de2018pq}, while for $n$-mode channels the theorem has been proven only for $p=1$ \cite{giovannetti2015majorization,holevo2015gaussian} and $p=q$ \cite{frank2017norms,holevo2017quantum}.
A proof of the Entropy Photon-number Inequality has been attempted with the quantum analogue of the heat semigroup technique of the proof of the Entropy Power Inequality by Blachman and Stam.
This technique led instead to the proof of the quantum Entropy Power Inequality \cite{konig2014entropy,konig2016corrections,konig2013limits,de2014generalization,de2015multimode,de2017gaussian} (Theorem \ref{thm:qEPI}), which provides a lower bound to the output entropy of the beam-splitter in terms of the entropies of the two inputs.
This bound is strictly lower than the output entropy achieved by Gaussian input states, hence the quantum Entropy Power Inequality is strictly weaker than the Entropy Photon-number Inequality, that is still an open conjecture.
The same heat semigroup technique led to the proof of the quantum conditional Entropy Power Inequality \cite{koenig2015conditional,de2018conditional} and of the quantum Entropy Power Inequality for the quantum additive noise channels both in the unconditioned \cite{huber2017geometric} and conditional \cite{huber2018conditional} versions.
The quantum conditional Entropy Power Inequality (Theorem \ref{thm:qcEPI}) determines the minimum quantum conditional von Neumann entropy of the output of the beam-splitter or of the squeezing among all the input states where the two inputs are conditionally independent given the memory and have given quantum conditional entropies.
This inequality has been exploited to prove an uncertainty relation for the conditional Wehrl entropy \cite{de2018uncertainty}.
These Entropy Power Inequalities have stimulated the proof of similar inequalities in different contexts, such as the qubit swap channel \cite{audenaert2016entropy,carlen2016quantum} and information combining \cite{hirche2018bounds}.
The implications among the main results and conjectures for quantum Gaussian channels are summarized in \autoref{fig:implications}.

As a possible approach towards the proof of the unsolved entropic inequalities for quantum Gaussian channels, we mention that sharp functional inequalities in the commutative setting have been recently studied using the theory of optimal transport \cite{villani_optimal_2009}.
These methods led to e.g. quantitative stability results for isoperimetric \cite{figalli_mass_2010}, Sobolev and log-Sobolev \cite{figalli_sharp_2013, fathi_quantitative_2016} inequalities.
Ideas from optimal transport are also implicit in the solution of Shannon's problem on the monotonicity of entropy \cite{artstein_solution_2004}.
Recently, transportation distances have been proposed in the quantum fermionic setting \cite{carlen_analog_2014, carlen_gradient_2017} and have then been extended to quantum Gaussian systems \cite{rouze2017concentration,rouze2017relating} (see also Ref. \onlinecite{datta2017contractivity}).

An interesting particular case of the inequalities for quantum Gaussian channels is when the input states are diagonal in the Fock basis \cite{guha2016thinning,johnson2017bruijn}.
This provides a link between quantum Gaussian channels and classical discrete probability theory.
The restriction of the one-mode quantum-limited attenuator to input states diagonal in the Fock basis is the linear map acting on discrete classical probability distributions on $\mathbb{N}$ known as thinning \cite{de2015passive}.
The quantum-limited attenuator is the quantum Gaussian channel that models the attenuation of electromagnetic signals.
The thinning has been introduced by R\'enyi \cite{renyi1956characterization} as a discrete analogue of the rescaling of a continuous real random variable, and has been involved with this role in discrete versions of the central limit theorem \cite{harremoes2007thinning,yu2009monotonic,harremoes2010thinning}, of the Entropy Power Inequality \cite{yu2009concavity,johnson2010monotonicity} and of Young's inequality \cite{lanconelli2013holder}. Most of these results require the ad hoc hypothesis of the ultra log-concavity (ULC) of the input state.
In particular, the Restricted Thinned Entropy Power Inequality \cite{johnson2010monotonicity} states that the Poisson input probability distribution minimizes the output Shannon entropy of the thinning among all the ULC input probability distributions with a given Shannon entropy.
The results on quantum Gaussian channels presented in this review led to the proof of new entropic inequalities for the thinning that apply to any probability distribution, regardless of whether they satisfy the ULC assumption.
Quantum Gaussian states correspond to geometric probability distributions.
The inequalities on quantum Gaussian channels imply that geometric input probability distributions both achieve the norms of the thinning \cite{de2018pq} (Theorem \ref{thm:pqT}) and minimize its output entropy among all the input probability distributions with a given entropy \cite{de2016gaussian} (Theorem \ref{thm:EPnIT}).

The review is structured as follows.
In \autoref{sec:GOFA}, we present the classical results for Gaussian optimizers in functional analysis.
In \autoref{sec:QGS}, we introduce Gaussian quantum systems and channels, and the necessary notions of quantum mechanics.
In \autoref{sec:MOE} we present the minimum output entropy conjecture and its proof.
In \autoref{sec:gaussopt}, we present all the conjectures on Gaussian optimizers in quantum information including the new sharp Young's inequality for the beam-splitter, together with the state of the art in their proofs.
In \autoref{sec:thin}, we present the thinning and its relation with the results for quantum Gaussian channel.
In \autoref{sec:EPI}, we present the quantum Entropy Power Inequality and its proof.
Moreover, we introduce the quantum conditional entropy, and present the conditioned version of the quantum Entropy Power Inequality.
We conclude in \autoref{sec:concl}.

\section{Gaussian optimizers in functional analysis}\label{sec:GOFA}
\subsection{Gaussian kernels have only Gaussian maximizers}
For any $p\ge1$, the $L^p(\mathbb{R}^n)$ norm of a function $f:\mathbb{R}^n\to\mathbb{C}$ is
\begin{equation}
\|f\|_p = \left(\int_{\mathbb{R}^n}{|f(x)|}^p\,\mathrm{d} x\right)^\frac{1}{p}\;.
\end{equation}
Given $p,\,q \ge 1$, let us consider a Gaussian integral kernel $G$ from $L^p(\mathbb{R}^m)$ to $L^q(\mathbb{R}^n)$:
\begin{equation}
(G\,f)(x) = \int_{\mathbb{R}^m} G(x,y)\,f(y)\,\mathrm{d}y\;,\qquad x\in\mathbb{R}^n\;,\qquad f\in L^p(\mathbb{R}^m)\;,
\end{equation}
where $G(x,y)$ is a Gaussian function on $\mathbb{R}^{m+n}$, i.e., the exponential of a quadratic polynomial.
The norm of $G$ is
\begin{equation}\label{eq:defpq}
\left\|G\right\|_{p\to q}=\sup_{0<\|f\|_p<\infty}\frac{\left\|G\,f\right\|_q}{\left\|f\right\|_p}\;.
\end{equation}
In the seminal paper ``Gaussian kernels have only Gaussian maximizers'' \cite{lieb1990Gaussian}, Lieb proved that under certain fairly broad assumptions on $G$, $p$ and $q$, this operator is well defined, and the supremum in \eqref{eq:defpq} is attained on a Gaussian function $f$.
If $1<p<q<\infty$, any function that attains the supremum in \eqref{eq:defpq} is a Gaussian function.
The proof of this fundamental result is based on the multiplicativity of the norm of generic integral kernels with respect to the tensor product.
\begin{thm}[Refs. \onlinecite{lieb1990Gaussian,holevo2006multiplicativity}]
The norms of integral kernels are multiplicative, i.e., for any two (not necessarily Gaussian) integral kernels $G_1:L^p(\mathbb{R}^{m_1})\to L^q(\mathbb{R}^{n_1})$ and $G_2:L^p(\mathbb{R}^{m_2})\to L^q(\mathbb{R}^{n_2})$,
\begin{equation}
  \|G_1\otimes G_2\|_{p\to q} = \|G_1\|_{p\to q}\|G_2\|_{p\to q}\;.
\end{equation}
Moreover, if the ratios $\left.\|G_1f_1\|_q\right/\|f_1\|_p$ and $\left.\|G_2f_2\|_q\right/\|f_2\|_p$ are maximized by the unique functions $f_1=\bar{f}_1\in L^p(\mathbb{R}^{m_1})$ and $f_2=\bar{f}_2\in L^p(\mathbb{R}^{m_2})$, the ratio $\left.\|(G_1\otimes G_2)f\|_q\right/\|f\|_p$ is maximized by the unique function $f=\bar{f}_1\otimes\bar{f}_2\in L^p(\mathbb{R}^{m_1+m_2})$.
\end{thm}

\subsection{The sharp Young's inequality for convolutions}
The convolution operation can be considered as a degenerate Gaussian integral kernel given by a Dirac delta function centered in the origin.
Indeed, the convolution of $f\in L^p(\mathbb{R}^n)$ with $g\in L^q(\mathbb{R}^n)$ is
\begin{equation}
(f*g)(x) = \int_{\mathbb{R}^{2n}}f(y)\,g(z)\,\delta_0(y+z-x)\,\mathrm{d}y\,\mathrm{d}z\;,\qquad x\in\mathbb{R}^n\;.
\end{equation}
The sharp Young's inequality for convolutions states that the supremum
\begin{equation}
\sup_{0<\|f\|_p,\|g\|_q<\infty}\frac{\|f*g\|_r}{\|f\|_p\|g\|_q}
\end{equation}
is finite iff
\begin{equation}
\frac{1}{p}+\frac{1}{q} = 1 + \frac{1}{r}\;,
\end{equation}
and in this case it is achieved by Gaussian functions.
This result has been first proven by Beckner \cite{beckner1975inequalities} and by Brascamp and Lieb \cite{brascamp2002best} using a rearrangement inequality for integrals \cite{brascamp2002general}.
A completely different proof based on the heat semigroup has been provided by Toscani \cite{toscani2014heat}.

\subsection{The Entropy Power Inequality}
Let $X$ be a random variable with values in $\mathbb{R}^n$ and whose probability law is absolutely continuous with respect to the Lebesgue measure, so that it admits a  probability density $f(x)\mathrm{d}x$.
The Shannon differential entropy \cite{cover2006elements} of $X$ is
\begin{equation}
S(X) = -\int_{\mathbb{R}^n}f(x)\ln f(x)\;\mathrm{d}x\;,
\end{equation}
and quantifies the noise contained in $X$.
Let $\sigma$ be a symmetric strictly positive $n\times n$ real matrix, and let $X$ be the centered Gaussian random variable with covariance matrix $\sigma$ and density
\begin{equation}
f(x) = \frac{\mathrm{e}^{-\frac{1}{2}x^T\sigma^{-1}x}}{\sqrt{\det(2\pi\sigma)}}\;.
\end{equation}
The Shannon differential entropy of $X$ is proportional to the logarithm of the determinant of the covariance matrix:
\begin{equation}\label{eq:SG}
S(X) = \frac{1}{2}\ln\det(2\pi\mathrm{e}\sigma)\;.
\end{equation}

Let us consider the sum of two independent random variables $X$ and $Y$ with values in $\mathbb{R}^n$.
The Entropy Power Inequality \cite{dembo1991information,stam1959some,shannon2001mathematical} states that, if $X$ and $Y$ have Shannon differential entropy fixed to the values $S(X)$ and $S(Y)$, respectively, the Shannon differential entropy of $X+Y$ is minimum when $X$ and $Y$ have a Gaussian probability distribution with proportional covariance matrices.
The covariance matrix of the sum of two independent random variables is equal to the sum of their covariance matrices:
\begin{equation}\label{eq:sum}
\sigma_{X+Y} = \sigma_X + \sigma_Y\;.
\end{equation}
If $\sigma_Y = \lambda\,\sigma_X$ for some $\lambda>0$, \eqref{eq:SG} and \eqref{eq:sum} imply
\begin{equation}
\exp\frac{2S(X+Y)}{n} = \exp\frac{2S(X)}{n} + \exp\frac{2S(Y)}{n}\;,
\end{equation}
so that the Entropy Power Inequality has the form
\begin{equation}\label{eq:cEPI}
\exp\frac{2S(X+Y)}{n} \ge \exp\frac{2S(X)}{n} + \exp\frac{2S(Y)}{n}\;.
\end{equation}
Two different proofs of the Entropy Power Inequality are known.
The first is due to Blachman and Stam \cite{stam1959some,blachman1965convolution}, and is based on perturbing the inputs $X$ and $Y$ with the heat semigroup.
The second is due to Lieb \cite{lieb1978proof}, and is based on the sharp Young's inequality for convolutions and on the properties of the R\'enyi entropies.
For any $p>1$, the $p$-R\'enyi entropy of the random variable $X$ with values in $\mathbb{R}^n$ and density $f$ is
\begin{equation}
S_p(X) = \frac{p}{1-p}\ln\|f\|_p\;.
\end{equation}
The R\'enyi entropies are a generalization of the Shannon differential entropy, which is recovered in the limit $p\to1$:
\begin{equation}
S(X) = \lim_{p\to 1}S_p(X)\;.
\end{equation}

\section{Quantum Gaussian systems}\label{sec:QGS}
In this Section, we introduce the elements of quantum information and quantum Gaussian systems that are needed for presenting the entropic inequalities for quantum Gaussian channels.
For a more comprehensive introduction, we refer the reader to the books Refs. \onlinecite{holevo2013quantum,serafini2017quantum} and the review Ref. \onlinecite{holevo2015gaussian}.

\subsection{Quantum systems}
Let $\mathcal{H}$ be a separable complex Hilbert space with not necessarily finite dimension.
We adopt the bra-ket notation, where a vector $\psi\in\mathcal{H}$ is denoted as $|\psi\rangle$ and the scalar product between the vectors $\phi$ and $\psi$ is denoted as $\langle\phi|\psi\rangle$, is linear in $\psi$ and antilinear in $\phi$.

For any $p\ge1$, the $p$-Schatten norm of a linear compact operator $\hat{X}$ on $\mathcal{H}$ is
\begin{equation}
\left\|\hat{X}\right\|_p = \left(\mathrm{Tr}\left(\hat{X}^\dag\hat{X}\right)^\frac{p}{2}\right)^\frac{1}{p}\;,
\end{equation}
where $\hat{X}^\dag$ is the adjoint operator of $\hat{X}$.
The $p$-Schatten norm play the role of the $L^p$ norm of functional analysis.
The operators with finite $1$-Schatten norm are called trace-class operators.
The $\infty$-Schatten norm $\|\hat{X}\|_\infty$ of a continuous linear operator $\hat{X}$ is defined as the supremum of the spectrum of $\sqrt{\hat{X}^\dag\hat{X}}$.

Quantum states are the noncommutative counterpart of probability measures.
A quantum state is a positive trace-class operator with unit trace.
Any quantum state $\hat{\rho}$ can be diagonalized in an orthonormal basis:
\begin{equation}\label{eq:rhodiag}
\hat{\rho} = \sum_{k=0}^\infty p_k\,|\psi_k\rangle\langle\psi_k|\;,
\end{equation}
where $\{|\psi_k\rangle\langle\psi_k|\}_{k\in\mathbb{N}}$ denote the rank-one projectors onto the orthonormal vectors $\{\psi_k\}_{k\in\mathbb{N}}$, and $\{p_k\}_{k\in\mathbb{N}}$ are the eigenvalues of $\hat{\rho}$.
Since $\hat{\rho}$ is positive and has unit trace, $\{p_k\}_{k\in\mathbb{N}}$ is a probability measure on $\mathbb{N}$.
The quantum state $\hat{\rho}$ is called \emph{pure} if it is a rank-one projector, and \emph{mixed} otherwise.
With a small abuse of nomenclature, we call pure state both the normalized vector $\psi\in\mathcal{H}$ and the associated rank-one projector $|\psi\rangle\langle\psi|$.
From \eqref{eq:rhodiag}, any quantum state can be expressed as a convex combination of orthogonal pure states.

The von Neumann entropy of the quantum state $\hat{\rho}$ in \eqref{eq:rhodiag} is the Shannon entropy of its eigenvalues:
\begin{equation}
S(\hat{\rho}) = -\mathrm{Tr}\left[\hat{\rho}\ln\hat{\rho}\right] = -\sum_{k=0}^\infty p_k\ln p_k\;,
\end{equation}
and is the quantum counterpart of the Shannon differential entropy.
As the Shannon entropy and contrarily to the Shannon differential entropy, the von Neumann entropy is always positive, and vanishes iff $\hat{\rho}$ is pure.
If $\hat{\rho}$ is a quantum state of the quantum system $A$ with Hilbert space $\mathcal{H}_A$, we use indistinctly the notations $S(\hat{\rho})$ or $S(A)$ for the entropy of $\hat{\rho}$.

As in the case of classical probability measures, we can define for any $p>1$ the $p$-R\'enyi entropy of the quantum state $\hat{\rho}$ as
\begin{equation}
S_p(\hat{\rho}) = \frac{p}{1-p}\ln\|\hat{\rho}\|_p\;.
\end{equation}
The R\'enyi entropies are a generalization of the von Neumann entropy, which is recovered in the limit $p\to 1$:
\begin{equation}\label{eq:limR}
S(\hat{\rho}) = \lim_{p\to 1}S_p(\hat{\rho})\;.
\end{equation}

The observables of a quantum system are the self-adjoint operators on the Hilbert space, and the expectation value of the observable $\hat{O}$ on the state $\hat{\rho}$ is
\begin{equation}
\left\langle\hat{O}\right\rangle_{\hat{\rho}} = \mathrm{Tr}\left[\hat{O}\,\hat{\rho}\right]\;.
\end{equation}

If $A$ and $B$ are quantum systems with Hilbert spaces $\mathcal{H}_A$ and $\mathcal{H}_B$, the joint system $AB$ has Hilbert space $\mathcal{H}_A\otimes\mathcal{H}_B$.
A pure state $\psi\in\mathcal{H}_A\otimes\mathcal{H}_B$ is called \emph{product state} if $\psi=\psi_A\otimes\psi_B$ for some $\psi_A\in\mathcal{H}_A$ and $\psi_B\in\mathcal{H}_B$.
A fundamental difference with respect to classical probability is that not all pure states are product states.
A pure state that is not a product state is called \emph{entangled state}.
Let $\hat{\rho}_{AB}$ be a quantum state of the joint quantum system $AB$.
We define the reduced or marginal states on $A$ and $B$ as
\begin{equation}
\hat{\rho}_A = \mathrm{Tr}_B\,\hat{\rho}_{AB}\;,\qquad \hat{\rho}_B = \mathrm{Tr}_A\,\hat{\rho}_{AB}\;,
\end{equation}
where $\mathrm{Tr}_A$ and $\mathrm{Tr}_B$ denote the partial trace over the system $A$ and $B$, respectively.
In other words, $\hat{\rho}_A$ is the quantum state of $A$ such that
\begin{equation}
\mathrm{Tr}_A\left[\hat{X}_A\,\hat{\rho}_A\right] = \mathrm{Tr}_{AB}\left[\left(\hat{X}_A\otimes\hat{\mathbb{I}}_B\right)\hat{\rho}_{AB}\right]
\end{equation}
for any bounded operator $\hat{X}_A$ on $\mathcal{H}_A$, and analogously for $\hat{\rho}_B$.

\subsection{Quantum channels}
Quantum channels \cite{nielsen2010quantum,hayashi2016quantum,wilde2017quantum,holevo2013quantum} are the noncommutative counterpart of the Markov operators of probability theory.
A quantum channel $\Phi$ from the quantum system $A$ with Hilbert space $\mathcal{H}_A$ to the quantum system $B$ with Hilbert space $\mathcal{H}_B$ is a linear completely positive trace-preserving map from the trace-class operators on $\mathcal{H}_A$ to the trace-class operators on $\mathcal{H}_B$.
Precisely, a map $\Phi$ is said to be
\begin{itemize}
\item positive if $\Phi(\hat{X})\ge0$ for any trace-class operator $\hat{X}\ge0$;
\item completely positive if the map $\Phi\otimes\mathbb{I}_d$ is positive for any $d\in\mathbb{N}$, where $\mathbb{I}_d$ denotes the identity map on the operators on the Hilbert space $\mathbb{C}^d$;
\item trace-preserving if $\mathrm{Tr}\,\Phi(\hat{X})=\mathrm{Tr}\,\hat{X}$ for any trace-class operator $\hat{X}$.
\end{itemize}
These properties ensure that for any $d\in\mathbb{N}$ the map $\Phi\otimes\mathbb{I}_d$ sends the quantum states on $\mathcal{H}_A\otimes\mathbb{C}^d$ to quantum states on $\mathcal{H}_B\otimes\mathbb{C}^d$.
Since any joint probability measure of two random variables is a convex combination of product probability measures, the complete positivity of any Markov operator is a trivial consequence of its positivity.
On the contrary, the existence of entanglement makes complete positivity a nontrivial requirement for quantum channels.
For example, the transposition is a linear positive trace-preserving map that is not completely positive.

Any quantum channel $\Phi$ from $A$ to $B$ can be realized by an isometry followed by a partial trace, i.e., there exists an Hilbert space $\mathcal{H}_E$ and an isometry $\hat{V}:\mathcal{H}_A\to\mathcal{H}_B\otimes\mathcal{H}_E$ with $\hat{V}^\dag\hat{V}=\hat{\mathbb{I}}_A$ such that for any trace-class operator $\hat{X}$ on $\mathcal{H}_A$,
\begin{equation}\label{eq:stinespring}
\Phi\left(\hat{X}\right) = \mathrm{Tr}_E\left[\hat{V}\,\hat{X}\,\hat{V}^\dag\right]\;.
\end{equation}
The Hilbert space $\mathcal{H}_E$ and the isometry $\hat{V}$ are unique up to isometries on $\mathcal{H}_E$.
The expression \eqref{eq:stinespring} is called the \emph{Stinespring dilation} of $\Phi$.
The quantum channel from $A$ to $E$ defined on trace-class operators as
\begin{equation}\label{eq:complementary}
\tilde{\Phi}\left(\hat{X}\right) = \mathrm{Tr}_A\left[\hat{V}\,\hat{X}\,\hat{V}^\dag\right]
\end{equation}
is called the \emph{complementary channel} of $\Phi$ \cite{holevo2013quantum}.
We mention that the Stinespring dilation has a classical analogue that has found some applications in the commutative setting, e.g., in infinite-dimensional stochastic analysis \cite{hausenblas2001note}.

Let $\Phi:A\to B$ be a quantum channel.
The \emph{dual channel} of $\Phi$ is the linear map $\Phi^\dag$ from bounded operators on $\mathcal{H}_B$ to bounded operators on $\mathcal{H}_A$ such that for any trace-class operator $\hat{A}$ on $\mathcal{H}_A$ and any bounded operator $\hat{B}$ on $\mathcal{H}_B$
\begin{equation}
\mathrm{Tr}\left[\hat{B}\,\Phi\left(\hat{A}\right)\right] = \mathrm{Tr}\left[\Phi^\dag\left(\hat{B}\right)\,\hat{A}\right]\;.
\end{equation}

For any $1\le p,q\le \infty$, the $p\to q$ norm of a quantum channel $\Phi$ is defined as
\begin{equation}
\|\Phi\|_{p\to q} = \sup_{0<\left\|\hat{X}\right\|_p<\infty}\frac{\left\|\Phi\left(\hat{X}\right)\right\|_q}{\left\|\hat{X}\right\|_p}\;.
\end{equation}
A fundamental question is whether the $p\to q$ norm of a channel is multiplicative with respect to the tensor product, i.e., whether
\begin{equation}
\left\|\Phi^{\otimes n}\right\|_{p\to q} = \|\Phi\|_{p\to q}^n\qquad \forall\;n\in\mathbb{N}\;.
\end{equation}
This property holds for any classical integral kernel \cite{lieb1990Gaussian,holevo2006multiplicativity}, but it is known to fail for generic quantum channels \cite{holevo2006multiplicativity}.

\subsection{Quantum Gaussian systems}
A $n$-mode Gaussian quantum system is the mathematical model for $n$ harmonic oscillators, or $n$ modes of the electromagnetic radiation.
For the sake of simplicity, we present one-mode Gaussian quantum systems first.

The Hilbert space of a one-mode Gaussian quantum system is $L^2(\mathbb{R})$, the irreducible representation of the canonical commutation relation (see Ref. \onlinecite{serafini2017quantum} or Ref. \onlinecite[Chapter 12]{holevo2013quantum} for a more complete presentation)
\begin{equation}
\left[\hat{a},\;\hat{a}^\dag\right]=\hat{\mathbb{I}}\;.
\end{equation}

The operator $\hat{a}$ is called ladder operator, plays the role of a noncommutative complex variable and acts on $\psi$ in a suitable dense domain in $L^2(\mathbb{R})$ as
\begin{equation}
(\hat{a}\,\psi)(x) = \frac{x\,\psi(x) + \psi'(x)}{\sqrt{2}}\;.
\end{equation}
The quantum states of a one-mode Gaussian quantum system are the quantum counterparts of the probability measures on $\mathbb{R}^2$. Since each mode is associated to a complex noncommutative variable, the number of real classical components is twice the number of quantum modes.
We define the Hamiltonian
\begin{equation}\label{eq:defH}
\hat{N}=\hat{a}^\dag\hat{a}\;,
\end{equation}
that counts the number of excitations, or photons.
The vector annihilated by $\hat{a}$ is the vacuum and is denoted by $|0\rangle$.
From the vacuum we can build the eigenstates of the Hamiltonian, called Fock states:
\begin{equation}\label{fock}
|n\rangle=\frac{\left(\hat{a}^\dag\right)^n}{\sqrt{n!}}|0\rangle\;,\quad\langle m|n\rangle=\delta_{mn}\;,\quad \hat{N}|n\rangle=n|n\rangle\;,\quad m,\,n\in\mathbb{N}\;,
\end{equation}
where $\langle\phi|\psi\rangle$ denotes the scalar product in $L^2(\mathbb{R})$.
An operator diagonal in the Fock basis is called Fock-diagonal.

A quantum Gaussian state is a quantum state proportional to the exponential of a quadratic polynomial in $\hat{a}$ and $\hat{a}^\dag$.
The most important Gaussian states are the thermal Gaussian states, where the polynomial is proportional to the Hamiltonian $\hat{a}^\dag\hat{a}$.
They correspond to a geometric probability distribution for the energy:
\begin{equation}\label{eq:omegaE}
\hat{\omega}(E)=\frac{1}{E+1}\sum_{n=0}^\infty \left(\frac{E}{E+1}\right)^n\;|n\rangle\langle n|\;,\qquad E\ge0\;.
\end{equation}
For $E=0$, we recover the vacuum state $\hat{\omega}(0)=|0\rangle\langle0|$.
The average energy of $\hat{\omega}(E)$ is
\begin{equation}
E=\mathrm{Tr}\left[\hat{N}\,\hat{\omega}(E)\right]\;,
\end{equation}
and the von Neumann entropy is
\begin{equation}\label{eq:defg}
g(E):=S(\hat{\omega}(E)) = \left(E+1\right)\ln\left(E+1\right)- E\ln E\;.
\end{equation}
As Gaussian probability measures maximize the Shannon differential entropy among all the probability measures with a given covariance matrix, thermal quantum Gaussian states maximize the von Neumann entropy among all the quantum states with a given average energy.

The Hilbert space of a $n$-mode Gaussian quantum system is the tensor product of $n$ Hilbert spaces of a one-mode Gaussian quantum system, i.e., the irreducible representation of the canonical commutation relations
\begin{equation}
\left[\hat{a}_i,\;\hat{a}^\dag_j\right]=\delta_{ij}\,\hat{\mathbb{I}}\;,\qquad \left[\hat{a}_i,\;\hat{a}_j\right]=\left[\hat{a}^\dag_i,\;\hat{a}^\dag_j\right]=0\;,\qquad i,\,j=1,\,\ldots,\,n\;,
\end{equation}
where each ladder operator $\hat{a}_i$ is associated to one mode.
An $n$-mode thermal quantum Gaussian state is the tensor product ${\hat{\omega}(E)}^{\otimes n}$ of $n$ identical one-mode thermal quantum Gaussian states.

\subsection{Quantum Gaussian channels}
Quantum Gaussian channels are the quantum channels that preserve the set of quantum Gaussian states.
The most important families of quantum Gaussian channels are the beam-splitter, the squeezing, the quantum Gaussian attenuators and the quantum Gaussian amplifiers.
The beam-splitter and the squeezing are the quantum counterparts of the classical linear mixing of random variables, and are the main transformations in quantum optics.
Let $A$ and $B$ be one-mode quantum Gaussian systems with ladder operators $\hat{a}$ and $\hat{b}$, respectively.
The \emph{beam-splitter} of transmissivity $0\le\lambda\le1$ is implemented by the unitary operator
\begin{equation}\label{defU}
\hat{U}_\lambda=\exp\left(\left(\hat{a}^\dag\hat{b}-\hat{b}^\dag\hat{a}\right)\arccos\sqrt{\lambda}\right)\;,
\end{equation}
and performs a linear rotation of the ladder operators (see e.g. Ref. \onlinecite[Section 1.4.2]{ferraro2005gaussian}):
\begin{align}\label{eq:defUlambda}
\hat{U}_\lambda^\dag\,\hat{a}\,\hat{U}_\lambda &= \sqrt{\lambda}\,\hat{a}+\sqrt{1-\lambda}\,\hat{b}\;,\nonumber\\
\hat{U}_\lambda^\dag\,\hat{b}\,\hat{U}_\lambda &= -\sqrt{1-\lambda}\,\hat{a}+\sqrt{\lambda}\,\hat{b}\;.
\end{align}
The physical beam-splitter is a passive element, and does not require energy for functioning.
Indeed, the mixing unitary operator preserves the Hamiltonian \eqref{eq:defH}:
\begin{equation}
\hat{U}_\lambda^\dag\left(\hat{a}^\dag\hat{a} + \hat{b}^\dag\hat{b}\right)\hat{U}_\lambda = \hat{a}^\dag\hat{a} + \hat{b}^\dag\hat{b}\;.
\end{equation}

The \emph{two-mode squeezing} \cite{barnett2002methods} of parameter $\kappa\ge1$ is implemented by the unitary operator
\begin{equation}\label{eq:defUk}
\hat{U}_\kappa=\exp\left(\left(\hat{a}^\dag\hat{b}^\dag-\hat{a}\,\hat{b}\right)\mathrm{arccosh}\sqrt{\kappa}\right)\;,
\end{equation}
and acts on the ladder operators as
\begin{align}
\hat{U}_\kappa^\dag\,\hat{a}\,\hat{U}_\kappa &= \sqrt{\kappa}\,\hat{a}+\sqrt{\kappa-1}\,\hat{b}^\dag\;,\nonumber\\
\hat{U}_\kappa^\dag\,\hat{b}\,\hat{U}_\kappa &= \sqrt{\kappa-1}\,\hat{a}^\dag+\sqrt{\kappa}\,\hat{b}\;.
\end{align}
The squeezing is an active operation that requires energy.
Indeed, the squeezing unitary operator does not preserve the Hamiltonian \eqref{eq:defH}.

We define for any joint quantum state $\hat{\rho}_{AB}$ on $AB$ and any $\lambda\ge0$ the quantum channel from $AB$ to $A$
\begin{equation}
\mathcal{B}_\lambda(\hat{\rho}_{AB}) = \mathrm{Tr}_B\left[\hat{U}_\lambda\,\hat{\rho}_{AB}\,\hat{U}_\lambda^\dag\right]\;,
\end{equation}
where $\mathrm{Tr}_B$ denotes the partial trace over the system $B$.
$\mathcal{B}_\lambda$ implements the beam-splitter for $0\le\lambda\le1$ and the squeezing for $\lambda\ge1$.

The quantum Gaussian attenuators model the attenuation and the noise affecting electromagnetic signals traveling through optical fibers or free space.
The \emph{quantum Gaussian attenuator} $\mathcal{E}_{\lambda,E}$ can be implemented mixing the input state $\hat{\rho}$ with the thermal Gaussian state with average energy $E\ge0$ through a beam-splitter of transmissivity $0\le\lambda\le1$:
\begin{equation}\label{eq:defE}
\mathcal{E}_{\lambda,E}(\hat{\rho}) = \mathcal{B}_\lambda(\hat{\rho}\otimes\hat{\omega}(E))\;.
\end{equation}
The quantum Gaussian attenuators constitute a multiplicative semigroup with composition law
\begin{equation}\label{eq:semE}
\mathcal{E}_{1,E}=\mathbb{I}\;,\qquad \mathcal{E}_{\lambda,E}\circ\mathcal{E}_{\lambda',E} = \mathcal{E}_{\lambda\lambda',E}\qquad \forall\;E\ge0,\;0\le\lambda,\lambda'\le1\;.
\end{equation}

The quantum Gaussian amplifiers model the amplification of electromagnetic signals.
The \emph{quantum Gaussian amplifier} $\mathcal{A}_{\kappa,E}$ can be implemented performing a two-mode squeezing of parameter $\kappa\ge1$ on the input state $\hat{\rho}$ and the thermal Gaussian state with average energy $E\ge0$:
\begin{equation}\label{eq:defA}
\mathcal{A}_{\kappa,E}(\hat{\rho}) = \mathcal{B}_\kappa(\hat{\rho}\otimes\hat{\omega}(E))\;.
\end{equation}
Also the quantum Gaussian amplifiers constitute a semigroup with composition law
\begin{equation}\label{eq:semA}
\mathcal{A}_{1,E}=\mathbb{I}\;,\qquad \mathcal{A}_{\kappa,E}\circ\mathcal{A}_{\kappa',E} = \mathcal{A}_{\kappa\kappa',E}\qquad \forall\;E\ge0,\;\kappa,\kappa'\ge1\;.
\end{equation}

The attenuator $\mathcal{E}_{\lambda,E}$ and the amplifier $\mathcal{A}_{\kappa,E}$ are called \emph{quantum-limited} if $E=0$, i.e., if they mix the input state with the vacuum.
Indeed, the vacuum as state of the environment adds the least possible noise to the input state.
In this case, since $\hat{\omega}(0)=|0\rangle\langle0|$ is a pure state, the expressions \eqref{eq:defE} and \eqref{eq:defA} are the Stinespring dilations of the corresponding channels.

\section{The minimum output entropy conjecture}\label{sec:MOE}
The minimum output entropy of a quantum channel plays a key role in the determination of its classical information capacity.
The following theorem has been a fundamental result in quantum communication theory.
\begin{thm}\label{thm:MOE}
For any $n\in\mathbb{N}$, the vacuum state minimizes the output entropy of the $n$-mode quantum Gaussian attenuators and of the $n$-mode quantum Gaussian amplifiers, i.e., for any $n$-mode quantum state $\hat{\rho}$, any $E\ge0$, $0\le\lambda\le1$ and $\kappa\ge1$
\begin{align}
S\left(\mathcal{E}^{\otimes n}_{\lambda,E}(\hat{\rho})\right) &\ge S\left(\mathcal{E}_{\lambda,E}^{\otimes n}\left({|0\rangle\langle0|}^{\otimes n}\right)\right) = n\,S(\mathcal{E}_{\lambda,E}(|0\rangle\langle 0|))\;,\nonumber\\
S\left(\mathcal{A}^{\otimes n}_{\kappa,E}(\hat{\rho})\right) &\ge S\left(\mathcal{A}_{\kappa,E}^{\otimes n}\left({|0\rangle\langle0|}^{\otimes n}\right)\right) = n\,S(\mathcal{A}_{\kappa,E}(|0\rangle\langle 0|))\;.
\end{align}
Therefore, the minimum output entropy of the quantum attenuators and amplifiers is additive.
\end{thm}
We stress that Theorem \ref{thm:MOE} is trivial for classical Gaussian channels, i.e., for Gaussian integral kernels that send probability measures on $\mathbb{R}^m$ to probability measures on $\mathbb{R}^n$.
Indeed, for the concavity of the entropy it is sufficient to prove Theorem \ref{thm:MOE} for pure input states.
In the classical case, the only pure probability measures are the Dirac delta functions, and they all achieve the same output entropy.
As we will see, the proof of Theorem \ref{thm:MOE} exploits tools of quantum information theory that do not have a classical counterpart: the complementary channel and the decomposition of any Gaussian channel as a quantum-limited attenuator followed by a quantum-limited amplifier.

The proof of Theorem \ref{thm:MOE} is based on majorization theory \cite{marshall2010inequalities}.
\begin{defn}[majorization]\label{def:maj}
We say that the quantum state $\hat{\rho}$ majorizes the quantum state $\hat{\sigma}$, and write $\hat{\rho}\succ\hat{\sigma}$, iff $\hat{\sigma}$ can be obtained applying to $\hat{\rho}$ a convex combination of unitary operators, i.e., iff there exists a probability measure $\mu$ on the set of unitary operators such that
\begin{equation}
\hat{\sigma} = \int\hat{U}\,\hat{\rho}\,\hat{U}^\dag\,\mathrm{d}\mu\left(\hat{U}\right)\;.
\end{equation}
\end{defn}
The link between majorization and the entropy is provided by the following property.
\begin{prop}
Let $\hat{\rho}$ and $\hat{\sigma}$ be quantum states such that $\hat{\rho}\succ\hat{\sigma}$.
Then, $f(\hat{\rho})\ge f(\hat{\sigma})$ for any unitarily invariant convex functional $f$ on the set of quantum states.
In particular,
\begin{itemize}
\item $\|\hat{\rho}\|_p \ge \|\hat{\sigma}\|_p$ for any $p\ge1$;
\item $S(\hat{\rho}) \le S(\hat{\sigma})$.
\end{itemize}
\end{prop}
Theorem \ref{thm:MOE} is a consequence of this more fundamental result.
\begin{thm}[majorization for quantum Gaussian channels]\label{thm:maj}
For any $n\in\mathbb{N}$ and for all the $n$-mode quantum Gaussian attenuators and amplifiers, the output generated by the vacuum input state majorizes the output generated by any other input state, i.e., for any $0\le\lambda\le1$, $\kappa\ge1$ and $E\ge0$ and for any $n$-mode quantum state $\hat{\rho}$,
\begin{equation}
\mathcal{E}_{\lambda,E}^{\otimes n}\left({|0\rangle\langle0|}^{\otimes n}\right) \succ \mathcal{E}_{\lambda,E}^{\otimes n}(\hat{\rho})\;,\qquad \mathcal{A}_{\kappa,E}^{\otimes n}\left({|0\rangle\langle0|}^{\otimes n}\right) \succ \mathcal{A}_{\kappa,E}^{\otimes n}(\hat{\rho})\;.
\end{equation}
\end{thm}
Besides Theorem \ref{thm:MOE}, a fundamental consequence of Theorem \ref{thm:maj} is the following.
\begin{cor}[$1\to p$ norms of quantum Gaussian channels]\label{cor:1p}
For any $p\ge1$ and any $n\in\mathbb{N}$, the vacuum input state achieves the $1\to p$ norm of the $n$-mode quantum Gaussian attenuators and amplifiers, i.e., for any $0\le\lambda\le1$, $\kappa\ge1$ and $E\ge0$
\begin{align}
\left\|\mathcal{E}_{\kappa,E}^{\otimes n}\right\|_{1\to p} &= \left\|\mathcal{E}^{\otimes n}_{\kappa,E}\left({|0\rangle\langle0|}^{\otimes n}\right)\right\|_p = \left\|\mathcal{E}_{\kappa,E}(|0\rangle\langle0|)\right\|_p^n\;,\nonumber\\
\left\|\mathcal{A}_{\kappa,E}^{\otimes n}\right\|_{1\to p} &= \left\|\mathcal{A}^{\otimes n}_{\kappa,E}\left({|0\rangle\langle0|}^{\otimes n}\right)\right\|_p = \left\|\mathcal{A}_{\kappa,E}(|0\rangle\langle0|)\right\|_p^n\;.
\end{align}
Therefore, the $1\to p$ norms of the quantum Gaussian attenuators and amplifiers are multiplicative.
\end{cor}

Theorem \ref{thm:MOE} was first proven by Giovannetti, Holevo and Garc\'ia-Patr\'on \cite{giovannetti2015solution}.
Shortly later, Mari, Giovannetti and Holevo realized that the same proof implies the more general Theorem \ref{thm:maj}, first for one-mode quantum Gaussian channels \cite{mari2014quantum}, and then for multi-mode quantum Gaussian channels \cite{giovannetti2015majorization,holevo2016proof}.
We present here a sketch of the proof.
For more details, the reader can also consult the review Ref. \onlinecite{holevo2015gaussian}.

The first step to prove Theorem \ref{thm:maj} is the following observation.
\begin{prop}\label{prop:dec}
For any $n$, any $n$-mode quantum Gaussian attenuator or amplifier can be decomposed as an $n$-mode quantum-limited attenuator followed by a $n$-mode quantum-limited amplifier.
\end{prop}
Theorem \ref{thm:maj} is trivial for the quantum-limited attenuator, since the vacuum is a fixed point.
Thanks to Proposition \ref{prop:dec}, it is sufficient to prove Theorem \ref{thm:maj} for the quantum-limited amplifier.
It is easy to see that it is sufficient to prove Theorem \ref{thm:maj} for pure input states.
The next step exploits the following properties:
\begin{prop}\label{prop:compl}
Let $\Phi$ be a quantum channel and $\tilde{\Phi}$ be its complementary channel.
Then, for any pure input state $\psi$, the quantum states $\Phi(|\psi\rangle\langle\psi|)$ and $\tilde{\Phi}(|\psi\rangle\langle\psi|)$ have the same spectrum.
\end{prop}
From Proposition \ref{prop:compl}, the optimal input states for $\Phi$ and $\tilde{\Phi}$ must coincide.
\begin{prop}\label{prop:complA}
The complementary channel of the quantum-limited amplifier is a quantum-limited attenuator followed by the same quantum-limited amplifier followed by the transposition, i.e., for any $\kappa\ge1$,
\begin{equation}
\tilde{\mathcal{A}}_{\kappa,0} = T\circ\mathcal{A}_{\kappa,0}\circ\mathcal{E}_{1-1/\kappa,0}\;,
\end{equation}
where $T$ is the transposition operation.
\end{prop}
From Propositions \ref{prop:compl} and \ref{prop:complA}, the optimal input states for the quantum-limited amplifier must coincide with the optimal input states for a suitable quantum-limited attenuator composed with the same quantum-limited amplifier.
Since the optimal input states must be pure, they must be left pure by the quantum-limited attenuator.
The claim then follows from the following property.
\begin{prop}
For any $n\in\mathbb{N}$ and any $0<\lambda<1$, the vacuum is the only $n$-mode quantum state $\hat{\rho}$ such that $\mathcal{E}_{\lambda,0}^{\otimes n}(\hat{\rho})$ is pure.
\end{prop}

\section{Gaussian optimizers for entropic inequalities in quantum information}\label{sec:gaussopt}
The problem of determining the information capacity region of the quantum Gaussian degraded broadcast channel has led to a constrained minimum output entropy conjecture \cite{guha2007classical}, which is a generalization of Theorem \ref{thm:MOE} with a constrained input entropy.
\begin{conj}[constrained minimum output entropy conjecture]\label{conj:MOE}
For any $n\in\mathbb{N}$, quantum Gaussian input states minimize the output entropy of the $n$-mode Gaussian quantum attenuators and amplifiers among all the input states with a given entropy.
In other words, let $\hat{\rho}$ be a generic $n$-mode quantum state, and let $\hat{\omega}$ be the one-mode thermal Gaussian state with entropy $S(\hat{\rho})/n$, so that $\hat{\omega}^{\otimes n}$ is the $n$-mode thermal Gaussian state with the same entropy as $\hat{\rho}$.
Then, for any $0\le\lambda\le1$, $\kappa\ge1$ and $E\ge0$,
\begin{align}
S\left(\mathcal{E}^{\otimes n}_{\lambda,E}(\hat{\rho})\right) &\ge S\left(\mathcal{E}^{\otimes n}_{\lambda,E}\left(\hat{\omega}^{\otimes n}\right)\right) = n\,S(\mathcal{E}_{\lambda,E}(\hat{\omega})) = n\,g\left(\lambda\,g^{-1}\left(\frac{S(\hat{\rho})}{n}\right) + \left(1-\lambda\right)E\right)\;,\nonumber\\
S\left(\mathcal{A}^{\otimes n}_{\kappa,E}(\hat{\rho})\right) &\ge S\left(\mathcal{A}^{\otimes n}_{\kappa,E}\left(\hat{\omega}^{\otimes n}\right)\right) = n\,S(\mathcal{A}_{\kappa,E}(\hat{\omega}))\nonumber\\
& = n\,g\left(\kappa\,g^{-1}\left(\frac{S(\hat{\rho})}{n}\right) + \left(\kappa-1\right)\left(E+1\right)\right)\;,
\end{align}
where the function $g$ has been defined in \eqref{eq:defg}.
\end{conj}
Conjecture \ref{conj:MOE} has been proven only in the one-mode case ($n=1$) by De Palma, Trevisan and Giovannetti \cite{de2016gaussian,de2016gaussiannew}, and has been extended to one-mode gauge-contravariant quantum Gaussian channels by Qi, Wilde and Guha \cite{qi2017minimum}.
The proof by De Palma et al. is based on the following fundamental majorization result for one-mode quantum Gaussian channels \cite{de2015passive}, which extends Theorem \ref{thm:maj}.
\begin{thm}\label{thm:maj2}
For any $0\le\lambda\le1$, $\kappa\ge1$ and $E\ge0$ and any one-mode quantum state $\hat{\rho}$,
\begin{equation}
\mathcal{E}_{\lambda,E}(\hat{\rho})\prec\mathcal{E}_{\lambda,E}\left(\hat{\rho}^\downarrow\right)\;,\qquad \mathcal{A}_{\kappa,E}(\hat{\rho})\prec\mathcal{A}_{\kappa,E}\left(\hat{\rho}^\downarrow\right)\;,
\end{equation}
where $\hat{\rho}^\downarrow$ is the passive rearrangement of $\hat{\rho}$, i.e., the passive state with the same spectrum as $\hat{\rho}$.
\end{thm}
We recall that a passive state is a quantum state that minimizes the average energy among all the quantum states with the same spectrum \cite{pusz1978passive,lenard1978thermodynamical,gorecki1980passive}.
If $\hat{\rho}$ is diagonalized in the orthonormal eigenbasis $\{\psi_n\}_{n\in\mathbb{N}}$ as
\begin{equation}
\hat{\rho} = \sum_{n=0}^\infty p_n\,|\psi_n\rangle\langle\psi_n|\;,\qquad p_0\ge p_1\ge\ldots\ge0\;,
\end{equation}
$\hat{\rho}^\downarrow$ is given by
\begin{equation}
\hat{\rho}^\downarrow = \sum_{n=0}^\infty p_n\,|n\rangle\langle n|\;,
\end{equation}
where $\{|n\rangle\}_{n\in\mathbb{N}}$ is the Fock basis.

From Theorem \ref{thm:maj2}, in the case of one mode the constrained minimization of the output entropy of Conjecture \ref{conj:MOE} can be restricted to passive input states.
Unfortunately, an analogue majorization theorem does not hold for more than one mode \cite{de2016passive}.

Conjecture \ref{conj:MOE} has first been proven for the one-mode quantum-limited attenuator \cite{de2016gaussian}.
The proof is based on the following isoperimetric inequality, that constitutes the infinitesimal version of the conjecture.
\begin{thm}[isoperimetric inequality for the one-mode quantum-limited attenuator]\label{thm:iso}
Among all the input states with a given entropy, quantum Gaussian input states maximize the derivative of the output entropy of the one-mode quantum-limited attenuator with respect to the attenuation parameter.
In other words, let $\hat{\rho}$ be a one-mode quantum state, and $\hat{\omega}$ the one-mode thermal Gaussian state with the same entropy as $\hat{\rho}$.
Then,
\begin{equation}\label{eq:iso}
\left.\frac{\mathrm{d}}{\mathrm{d}\lambda}S\left(\mathcal{E}_{\lambda,0}(\hat{\rho})\right)\right|_{\lambda=1} \le \left.\frac{\mathrm{d}}{\mathrm{d}\lambda}S\left(\mathcal{E}_{\lambda,0}(\hat{\omega})\right)\right|_{\lambda=1} = g^{-1}(S(\hat{\rho}))\;g'\left(g^{-1}(S(\hat{\rho}))\right)\;.
\end{equation}
\end{thm}

The adjective ``isoperimetric'' is due to the formal analogy between entropy and volume \cite{dembo1991information}. Up to a change of signs, the left hand side in \eqref{eq:iso} plays the role of a perimeter and the function $g^{-1}(s)g'(g^{-1}(s))$ that of an isoperimetric profile.

Thanks to Theorem \ref{thm:maj2}, it is sufficient to prove Theorem \ref{thm:iso} for passive states.
The proof is then performed through the Lagrange multipliers.
Since the Hilbert space of a one-mode Gaussian quantum system has infinite dimension, a generic passive state has infinite parameters.
This issue is solved restricting to a finite dimensional subspace with bounded maximum energy, and then proving that the maximum of the left-hand side of \eqref{eq:iso} for passive input states supported in the subspace tends to the right-hand side in the limit of infinite maximum energy.

Conjecture \ref{conj:MOE} for the one-mode quantum-limited attenuator then follows integrating the isoperimetric inequality \eqref{eq:iso} thanks to the semigroup property \eqref{eq:semE} of the quantum-limited attenuator.

The generalization of Theorem \ref{thm:iso} to all the one-mode quantum Gaussian attenuators and amplifiers would have implied Conjecture \ref{conj:MOE} for $n=1$.
However, for any one-mode quantum Gaussian channel other than the quantum-limited attenuator, the infinite dimension of the Hilbert space is really an issue.
Indeed, for any quantum state $\hat{\rho}$ with a support of finite dimension $\left.\frac{\mathrm{d}}{\mathrm{d}\lambda}S\left(\mathcal{E}_{\lambda,E}(\hat{\rho})\right)\right|_{\lambda=1}$ is infinite for any $E>0$ and $\left.\frac{\mathrm{d}}{\mathrm{d}\kappa}S\left(\mathcal{A}_{\lambda,E}(\hat{\rho})\right)\right|_{\kappa=1}$ is infinite for any $E\ge0$, and nothing can be proven restricting to a finite dimensional subspace.
If one tries to use the Lagrange multipliers directly for the infinite dimensional problem, the Gaussian state is not the only solution \cite{qi2017minimum}, so that a new approach is needed.
This approach is based on the $p\to q$ norms and is presented in \autoref{ss:pq} below.

\subsection{Quantum Gaussian channels have Gaussian maximizers}\label{ss:pq}
The theorem ``Gaussian kernels have Gaussian maximizers'' has been conjectured to apply also to quantum Gaussian channels.
\begin{conj}[quantum Gaussian channels have Gaussian maximizers]\label{conj:pq}
For any $n\in\mathbb{N}$ and any $p,\,q\ge1$, quantum Gaussian input states achieve the $p\to q$ norm of the $n$-mode Gaussian quantum attenuators and amplifiers.
In other words, for any $0\le\lambda\le1$, $\kappa\ge1$ and $E\ge0$,
\begin{align}\label{eq:pq}
\left\|\mathcal{E}_{\lambda,E}^{\otimes n}\right\|_{p\to q} &= \sup_{E'\ge0}\frac{\left\|\mathcal{E}_{\lambda,E}^{\otimes n}\left({\hat{\omega}(E')}^{\otimes n}\right)\right\|_q}{\left\|{\hat{\omega}(E')}^{\otimes n}\right\|_p} = \left(\sup_{E'\ge0}\frac{\left\|\mathcal{E}_{\lambda,E}\left({\hat{\omega}(E')}\right)\right\|_q}{\left\|{\hat{\omega}(E')}\right\|_p}\right)^n\;,\nonumber\\
\left\|\mathcal{A}_{\kappa,E}^{\otimes n}\right\|_{p\to q} &= \sup_{E'\ge0}\frac{\left\|\mathcal{A}_{\kappa,E}^{\otimes n}\left({\hat{\omega}(E')}^{\otimes n}\right)\right\|_q}{\left\|{\hat{\omega}(E')}^{\otimes n}\right\|_p} = \left(\sup_{E'\ge0}\frac{\left\|\mathcal{A}_{\kappa,E}\left({\hat{\omega}(E')}\right)\right\|_q}{\left\|{\hat{\omega}(E')}\right\|_p}\right)^n\;,
\end{align}
where $\hat{\omega}(E')$ is the one-mode thermal Gaussian state with average energy $E'$ as in \eqref{eq:omegaE}.
Therefore, the $p\to q$ norms of the quantum Gaussian attenuators and amplifiers are multiplicative.
\end{conj}
\begin{rem}
The suprema in \eqref{eq:pq} are
\begin{itemize}
\item finite and achieved for a finite $E'\ge0$ if $1\le p < q$;
\item finite and asymptotically achieved in the limit $E'\to\infty$ if $1<p=q$;
\item infinite and asymptotically achieved in the limit $E'\to\infty$ if $1\le q<p$.
\end{itemize}
\end{rem}
\begin{rem}
Conjecture \ref{conj:pq} can be extended to any linear and completely positive map that preserves the set of unnormalized quantum Gaussian states, i.e., the operators proportional to a quantum Gaussian state.
These maps include all quantum Gaussian channels and all the probabilistic maps resulting from the conditioning on the outcome of a Gaussian measurement performed on a subsystem \cite{holevo2013quantum,serafini2017quantum}.
The generalized conjecture states that quantum Gaussian input states achieve the $p\to q$ norms of all such maps.
In this more general setup, the analogue of the optimization in the right-hand side of \eqref{eq:pq} cannot be restricted to the thermal Gaussian states, but has to be performed over all quantum Gaussian states.
\end{rem}
Conjecture \ref{conj:pq} has been proven only in some particular cases.
As we have seen in Corollary \ref{cor:1p}, the majorization result Theorem \ref{thm:maj} implies Conjecture \ref{conj:pq} for any $n$ in the case $p=1$.
De Palma, Trevisan and Giovannetti proved Conjecture \ref{conj:pq} in the case of one-mode quantum-limited channels, i.e., $n=1$ and $E=0$ \cite{de2018pq}.
Frank and Lieb proved Conjecture \ref{conj:pq} for any $n$ in the case $p=q$ \cite{frank2017norms}, and Holevo extended the result to any $n$-mode quantum Gaussian channel (still for $p=q$) \cite{holevo2017quantum}.

\subsubsection{The proof of Conjecture \ref{conj:pq} for one-mode quantum-limited Gaussian channels}
First, De Palma et al. prove Conjecture \ref{conj:pq} for the one-mode quantum-limited attenuator.
From the following Lemma, it is sufficient to prove Conjecture \ref{conj:pq} for positive input operators.
\begin{lem}[Ref. \onlinecite{audenaert2009note}]\label{lem:pos}
For any $p\ge1$, any quantum channel $\Phi$ and any operator $\hat{X}$,
\begin{equation}
\left\|\Phi\left(\hat{X}\right)\right\|_p \le \left\|\Phi\left(\sqrt{\hat{X}^\dag\hat{X}}\right)\right\|_p\;.
\end{equation}
\end{lem}
The proof of Conjecture \ref{conj:pq} is then based on the following new logarithmic Sobolev inequality, that constitutes the infinitesimal version of Conjecture \ref{conj:pq} (in the same way as Gross' logarithmic Sobolev inequality  is the infinitesimal version of Nelson's Hypercontractive theorem \cite{gross1993logarithmic}).

\begin{thm}[logarithmic Sobolev inequality for the quantum-limited Gaussian attenuator]\label{thm:logs}
Let us fix $p\ge1$.
Let $\hat{\rho}$ be a one-mode quantum state, and let $\hat{\omega}$ be the thermal Gaussian state such that $\hat{\omega}^p/\mathrm{Tr}\hat{\omega}^p$ has the same entropy as $\hat{\rho}^p/\mathrm{Tr}\hat{\rho}^p$.
Then,
\begin{equation}\label{eq:logs}
\left.\frac{\mathrm{d}}{\mathrm{d}\lambda}\ln\left\|\mathcal{E}_{\lambda,0}(\hat{\rho})\right\|_p\right|_{\lambda=1} \ge \left.\frac{\mathrm{d}}{\mathrm{d}\lambda}\ln\left\|\mathcal{E}_{\lambda,0}(\hat{\omega})\right\|_p\right|_{\lambda=1}\;.
\end{equation}
\end{thm}
Thanks to Theorem \ref{thm:maj2}, it is sufficient to prove Theorem \ref{thm:logs} for passive input states.
As in the case of Theorem \ref{thm:iso}, the proof is then performed through the Lagrange multipliers, restricting to a finite dimensional subspace with bounded maximum energy.
Conjecture \ref{conj:pq} for the one-mode quantum-limited attenuator follows integrating \eqref{eq:logs} thanks to the semigroup property of the attenuator \eqref{eq:semE}.

Conjecture \ref{conj:pq} for the one-mode quantum-limited amplifier follows from the following duality Lemma for the Schatten norms.
\begin{lem}\label{lem:duality}
For any $p>1$ and any positive operator $\hat{X}$,
\begin{equation}
\left\|\hat{X}\right\|_p = \sup\left\{\mathrm{Tr}\left[\hat{X}\,\hat{Y}\right]\left/\left\|\hat{Y}\right\|_\frac{p}{p-1}\right.:\hat{Y}\ge0,\;\mathrm{rank}\, \hat{Y}<\infty\right\}\;.
\end{equation}
\end{lem}
Lemma \ref{lem:duality} implies the following duality for the norms of quantum channels.
\begin{lem}
For any quantum channel $\Phi$ and any $p,\,q\ge1$,
\begin{equation}
\|\Phi\|_{p\to q} = \left\|\Phi^\dag\right\|_{\frac{q}{q-1}\to\frac{p}{p-1}}\;.
\end{equation}
\end{lem}
The norms of the quantum-limited amplifier can then be determined from the norms of the quantum-limited attenuator thanks to the following property.
\begin{lem}
The dual of the quantum-limited Gaussian amplifier is proportional to a quantum-limited Gaussian attenuator, i.e., for any $\kappa\ge1$,
\begin{equation}
\mathcal{A}_{\kappa,0}^\dag = \frac{1}{\kappa}\mathcal{E}_{\frac{1}{\kappa},0}\;.
\end{equation}
\end{lem}

\subsubsection{The proof of Conjecture \ref{conj:MOE} for all the one-mode attenuators and amplifiers}\label{sss:pqS}
De Palma, Trevisan and Giovannetti have exploited the proof of Conjecture \ref{conj:pq} for the one-mode quantum-limited amplifier to prove Conjecture \ref{conj:MOE} for all the one-mode attenuators and amplifiers \cite{de2016gaussiannew}.
First, they prove Conjecture \ref{conj:MOE} for the one-mode quantum-limited amplifier.
The first step is rephrasing Conjecture \ref{conj:pq} for the one-mode quantum-limited amplifier in the following way.
\begin{thm}
Let us fix $\kappa\ge1$.
Let $\hat{\rho}$ be a generic one-mode quantum state, and let $\hat{\omega}$ be the one-mode thermal Gaussian state with the same entropy as $\hat{\rho}$.
Then, for any $q>1$ there exists $1\le p<q$ such that the $p\to q$ norm of $\mathcal{A}_{\kappa,0}$ is achieved by $\hat{\omega}$, and
\begin{equation}\label{eq:pqA}
\frac{\|\mathcal{A}_{\kappa,0}(\hat{\rho})\|_q}{\|\hat{\rho}\|_p} \le \|\mathcal{A}_{\kappa,0}\|_{p\to q} = \frac{\|\mathcal{A}_{\kappa,0}(\hat{\omega})\|_q}{\|\hat{\omega}\|_p}\;.
\end{equation}
\end{thm}
Rewriting \eqref{eq:pqA} in terms of the R\'enyi entropies we get
\begin{equation}
S_q(\mathcal{A}_{\kappa,0}(\hat{\rho})) \ge S_q(\mathcal{A}_{\kappa,0}(\hat{\omega})) + \frac{p-1}{q-1}\frac{q}{p}\left(S_p(\hat{\rho}) - S_p(\hat{\omega})\right)\;.
\end{equation}
Taking the limit $q\to 1$ and recalling \eqref{eq:limR} we get the claim
\begin{equation}
S(\mathcal{A}_{\kappa,0}(\hat{\rho})) \ge S(\mathcal{A}_{\kappa,0}(\hat{\omega}))\;.
\end{equation}
This result implies Conjecture \ref{conj:MOE} for all the one-mode attenuators and amplifiers since any of these channels can be decomposed as a one-mode quantum-limited attenuator followed by a one-mode quantum-limited amplifier (Proposition \ref{prop:dec}), for which Conjecture \ref{conj:MOE} holds.

\subsubsection{The proof of Conjecture \ref{conj:pq} for p=q}
The proof by Frank and Lieb is completely different from the proof by De Palma et al., and is based on the following theorem.
\begin{thm}\label{thm:infty}
For any $p>1$ and any quantum channel $\Phi$,
\begin{equation}
\|\Phi\|_{p\to p} \le \left\|\Phi\left(\hat{\mathbb{I}}\right)\right\|_{\infty}^\frac{p-1}{p}\left\|\Phi^\dag\left(\hat{\mathbb{I}}\right)\right\|^\frac{1}{p}_{\infty}\;.
\end{equation}
\end{thm}
Conjecture \ref{conj:pq} follows directly applying Theorem \ref{thm:infty} to quantum Gaussian channels.
The proof of Theorem \ref{thm:infty} is based on Hadamard's three line lemma \cite{simon2015basic}.
\begin{thm}[Hadamard's three line lemma]\label{thm:had}
Let $f$ be analytic in the strip $\{z:0<\Re z<1\}$ and continuous and bounded on its closure.
Let
\begin{equation}
M_t(f) = \sup_{y\in\mathbb{R}} |f(t+\mathrm{i} y)|
\end{equation}
for $0\le t\le1$.
Then
\begin{equation}
M_t \le M_0^{1-t} M_1^t\;.
\end{equation}
\end{thm}
Theorem \ref{thm:infty} follows applying Theorem \ref{thm:had} to
\begin{equation}
f(z) = \mathrm{Tr}\left[\hat{Y}^{p\frac{1-z}{p-1}}\Phi\left(\hat{X}^{pz}\right)\right]\;,
\end{equation}
where $\hat{X}$ and $\hat{Y}$ are positive and $\hat{Y}$ has finite rank, and recalling the duality relation for the Schatten norms (Lemma \ref{lem:duality}).

\subsection{The Entropy Photon-number Inequality}
The Entropy Photon-number Inequality is the quantum counterpart of the Entropy Power Inequality for the beam-splitter.
Guha, Erkmen and Shapiro conjectured it \cite{guha2008entropy} as a generalization of Conjecture \ref{conj:MOE}, and De Palma, Mari and Giovannetti extended the conjecture to the squeezing \cite{de2014generalization}.
\begin{conj}[Entropy Photon-number Inequality]\label{conj:EPnI}
For any $n$, $n$-mode thermal quantum Gaussian states minimize the output entropy of the $n$-mode beam-splitter or squeezing among all the $n$-mode input states where the two inputs have given entropies.
In other words, let $\hat{\rho}_A$ and $\hat{\rho}_B$ be two $n$-mode quantum states, and let $\hat{\omega}_A$ and $\hat{\omega}_B$ be the one-mode thermal Gaussian states with entropies $S(\hat{\rho}_A)/n$ and $S(\hat{\rho}_B)/n$, such that $\hat{\omega}_A^{\otimes n}$ and $\hat{\omega}_B^{\otimes n}$ are the $n$-mode thermal Gaussian states with the same entropy as $\hat{\rho}_A$ and $\hat{\rho}_B$, respectively.
Then, for any $0\le\lambda\le 1$
\begin{align}\label{eq:EPnI}
S\left(\mathcal{B}_\lambda^{\otimes n}(\hat{\rho}_A\otimes\hat{\rho}_B)\right) &\ge S\left(\mathcal{B}_\lambda^{\otimes n}\left(\hat{\omega}_A^{\otimes n}\otimes\hat{\omega}_B^{\otimes n}\right)\right)\nonumber\\
&= n\;g\left(\lambda\;g^{-1}\left(\frac{S(\hat{\rho}_A)}{n}\right) + \left(1-\lambda\right)g^{-1}\left(\frac{S(\hat{\rho}_B)}{n}\right)\right)\;,
\end{align}
and for any $\kappa\ge1$
\begin{align}
S\left(\mathcal{B}_\kappa^{\otimes n}(\hat{\rho}_A\otimes\hat{\rho}_B)\right) &\ge S\left(\mathcal{B}_\kappa^{\otimes n}\left(\hat{\omega}_A^{\otimes n}\otimes\hat{\omega}_B^{\otimes n}\right)\right)\nonumber\\
&= n\;g\left(\kappa\;g^{-1}\left(\frac{S(\hat{\rho}_A)}{n}\right) + \left(\kappa-1\right)\left(g^{-1}\left(\frac{S(\hat{\rho}_B)}{n}\right)+1\right)\right)\;,
\end{align}
where the function $g$ has been defined in \eqref{eq:defg}.
\end{conj}
For any $n$-mode quantum state $\hat{\rho}$, the $n$-mode thermal Gaussian state with the same entropy as $\hat{\rho}$ has average photon number per mode $g^{-1}(S(\hat{\rho}/n))$.
This quantity is called the \emph{entropy photon-number} of $\hat{\rho}$, hence the name Entropy Photon-number Inequality.

In the case where the second input $\hat{\rho}_B$ of the beam-splitter or of the squeezing is a thermal Gaussian state, Conjecture \ref{conj:EPnI} reduces to Conjecture \ref{conj:MOE}.
The only other particular case where the Entropy Photon-number Inequality has been proven is when the two inputs are (not necessarily thermal) Gaussian states \cite{de2017gaussian}.

\subsection{The sharp Young's inequality for the beam-splitter}
The similarity between the Entropy Photon-number Inequality and the Entropy Power Inequality together with the proof of the latter through the sharp Young's inequality for convolutions leads to conjecture a quantum version of the Young's inequality, here formulated for the first time.

Let us define for any $n\in\mathbb{N}$, any $p,\,q,\,r\ge1$ and any $\lambda\ge0$
\begin{equation}\label{eq:defC}
C_n(p,q,r,\lambda) = \sup_{0<\|\hat{X}\|_p,\|\hat{Y}\|_q<\infty}\frac{\left\|\mathcal{B}_\lambda^{\otimes n}\left(\hat{X}\otimes \hat{Y}\right)\right\|_r}{\left\|\hat{X}\right\|_p\left\|\hat{Y}\right\|_q}\;.
\end{equation}

\begin{conj}[Quantum sharp Young's inequality]\label{conj:young}
For any $n\in\mathbb{N}$, any $p,\,q,\,r\ge1$ and any $\lambda\ge0$, the supremum in \eqref{eq:defC} can be restricted to thermal Gaussian states, i.e.,
\begin{align}\label{eq:young}
C_n(p,q,r,\lambda) &= \sup_{E_A,\,E_B\ge0}\frac{\left\|\mathcal{B}_\lambda^{\otimes n}\left({\hat{\omega}(E_A)}^{\otimes n}\otimes {\hat{\omega}(E_B)}^{\otimes n}\right)\right\|_r}{\left\|{\hat{\omega}(E_A)}^{\otimes n}\right\|_p\left\|{\hat{\omega}(E_B)}^{\otimes n}\right\|_q}\nonumber\\
&= \left(\sup_{E_A,\,E_B\ge0}\frac{\left\|\mathcal{B}_\lambda\left(\hat{\omega}(E_A)\otimes \hat{\omega}(E_B)\right)\right\|_r}{\left\|\hat{\omega}(E_A)\right\|_p\left\|\hat{\omega}(E_B)\right\|_q}\right)^n = {C_1(p,q,r,\lambda)}^n\;.
\end{align}
Therefore, the constants $C_n$ are multiplicative.
\end{conj}
\begin{rem}
We conjecture that the supremum in \eqref{eq:young} is
\begin{itemize}
\item finite and achieved by finite $E_A,\,E_B\ge0$ if $\frac{1}{p}+\frac{1}{q}>1+\frac{1}{r}$;
\item finite and asymptotically achieved in the limit $E_A,\,E_B\to\infty$ if $\frac{1}{p}+\frac{1}{q}=1+\frac{1}{r}$;
\item infinite and asymptotically achieved in the limit $E_A,\,E_B\to\infty$ if $\frac{1}{p}+\frac{1}{q}<1+\frac{1}{r}$.
\end{itemize}
The striking difference with respect to the classical case is that the supremum in \eqref{eq:young} is finite when $\frac{1}{p}+\frac{1}{q}>1+\frac{1}{r}$.
The divergence in the classical Young's inequality when $\frac{1}{p}+\frac{1}{q}>1+\frac{1}{r}$ is asymptotically achieved by a sequence of Gaussian probability measures that tends to a Dirac delta, and can be ascribed to the fact that a probability density can have arbitrarily high $L^\infty$ norm.
The divergence disappears in the quantum scenario since $\|\hat{\rho}\|_\infty\le1$ for any quantum state $\hat{\rho}$.
\end{rem}

The quantum sharp Young's inequality provides a multiplicative upper bound to the $p\to q$ norms of the quantum Gaussian attenuators and amplifiers.
Indeed, assuming Conjecture \ref{conj:young}, we have for any $n\in\mathbb{N}$, $p,\,q\ge1$, $0\le\lambda\le1$ and $E\ge0$
\begin{align}\label{eq:pqbound}
\left\|\mathcal{E}^{\otimes n}_{\lambda,E}\right\|_{p\to q} &= \sup_{0<\|\hat{X}\|_p<\infty}\frac{\left\|\mathcal{E}_{\lambda,E}^{\otimes n}\left(\hat{X}\right)\right\|_q}{\left\|\hat{X}\right\|_p} = \sup_{0<\|\hat{X}\|_p<\infty}\frac{\left\|\mathcal{B}_\lambda^{\otimes n}\left(\hat{X}\otimes{\hat{\omega}(E)}^{\otimes n}\right)\right\|_q}{\left\|\hat{X}\right\|_p}\nonumber\\
&\le\left(\inf_{r\ge1}C_1(p,r,q,\lambda)\left\|\hat{\omega}(E)\right\|_r\right)^n\;,
\end{align}
and the same holds for the Gaussian quantum amplifiers.
Since the conjectured quantum sharp Young's inequality is saturated by quantum Gaussian states, we conjecture that the upper bound \eqref{eq:pqbound} is sharp and coincides with \eqref{eq:pq}, i.e., that
\begin{equation}
\sup_{E'\ge0}\frac{\left\|\mathcal{E}_{\lambda,E}\left({\hat{\omega}(E')}\right)\right\|_q}{\left\|{\hat{\omega}(E')}\right\|_p} = \inf_{r\ge1}C_1(p,r,q,\lambda)\left\|\hat{\omega}(E)\right\|_r\;.
\end{equation}
Moreover, the quantum sharp Young's inequality provides a lower bound to the output entropy of the beam-splitter and of the squeezing.
Indeed, rewriting \eqref{eq:young} in terms of the R\'enyi entropies we get for any $n\in\mathbb{N}$, $\lambda\ge0$, $p,\,q,\,r\ge1$ and any $n$-mode quantum states $\hat{\rho}_A$ and $\hat{\rho}_B$
\begin{equation}\label{eq:SrC}
S_r\left(\mathcal{B}_\lambda^{\otimes n}\left(\hat{\rho}_A\otimes\hat{\rho}_B\right)\right) \ge \frac{r}{r-1}\left(\frac{p-1}{p}S_p(\hat{\rho}_A) + \frac{q-1}{q}S_q(\hat{\rho}_B) - n\ln C_1(p,q,r,\lambda)\right)\;.
\end{equation}
We choose $0\le\alpha,\beta<\frac{r}{r-1}$ and set
\begin{equation}
p = p(r,\alpha) = \frac{r}{r+\alpha-\alpha\,r}\;,\qquad q = q(r,\beta) = \frac{r}{r+\beta-\beta\,r}\;,
\end{equation}
so that \eqref{eq:SrC} becomes
\begin{equation}
S_r\left(\mathcal{B}_\lambda^{\otimes n}\left(\hat{\rho}_A\otimes\hat{\rho}_B\right)\right) \ge \alpha\,S_{p(r,\alpha)}(\hat{\rho}_A) + \beta\,S_{q(r,\beta)}(\hat{\rho}_B) - \frac{n\,r}{r-1}\ln C_1(p(r,\alpha),\,q(r,\beta),\,r,\,\lambda)\;.
\end{equation}
Finally, taking the limit $r\to1$ and the supremum over $\alpha,\beta\ge0$ we get
\begin{equation}\label{eq:EPnIbound}
S\left(\mathcal{B}_\lambda^{\otimes n}\left(\hat{\rho}_A\otimes\hat{\rho}_B\right)\right) \ge \sup_{\alpha,\beta\ge0}\left( \alpha\,S(\hat{\rho}_A) + \beta\,S(\hat{\rho}_B) - n\left.\frac{\mathrm{d}}{\mathrm{d}r}C_1(p(r,\alpha),\,q(r,\beta),\,r,\,\lambda)\right|_{r=1}\right)\;,
\end{equation}
where we used that
\begin{equation}
\lim_{r\to1}p(r,\alpha) = \lim_{r\to1}q(r,\beta) = 1\;,\qquad C_1(1,1,1,\lambda) = 1\;.
\end{equation}
Since the conjectured quantum Young inequality is saturated by quantum Gaussian states, we conjecture that the lower bound \eqref{eq:EPnIbound} is sharp and coincides with the bound provided by the Entropy Photon-number Inequality \eqref{eq:EPnI}.

\section{The thinning}\label{sec:thin}
The thinning \cite{renyi1956characterization} is the map acting on probability distributions on $\mathbb{N}$ that is the discrete analogue of the continuous rescaling operation on $\mathbb{R}_+$.
\begin{defn}[Thinning]
Let $N$ be a random variable with values in $\mathbb{N}$.
The thinning with parameter $0\leq\lambda\leq1$ is defined as
\begin{equation}
T_\lambda(N)=\sum_{i=1}^N B_i\;,
\end{equation}
where the $\{B_n\}_{n\in\mathbb{N}^+}$ are independent Bernoulli variables with parameter $\lambda$ (also independent of $N$), i.e., each $B_i$ is $1$ with probability $\lambda$, and $0$ with probability $1-\lambda$.
\end{defn}
From a physical point of view, the thinning can be understood as follows.
Let $p$ be the probability distribution of the number $N$ of photons that are sent through a beam-splitter with transmissivity $\lambda$, such that for any $n\in\mathbb{N}$ the probability that $n$ photons are sent is $p_n$.
Each photon has probability $\lambda$ of being transmitted and probability $1-\lambda$ of being reflected.
Then, $T_\lambda(N)$ is the random variable associated to the number of transmitted photons, and has probability distribution
\begin{equation}\label{Tn}
\left[T_\lambda(p)\right]_n=\sum_{k=n}^\infty \binom{k}{n}\lambda^n(1-\lambda)^{k-n}\;p_k\qquad \forall\;n\in\mathbb{N}\;.
\end{equation}

The thinning coincides with the restriction of the one-mode quantum-limited Gaussian attenuator to input states diagonal in the Fock basis.
\begin{thm}[Ref. {\onlinecite[Theorem 56]{de2015passive}}]\label{thm:thinatt}
For any $0\le\lambda\le1$ and any probability distribution $p$ on $\mathbb{N}$
\begin{equation}
\mathcal{E}_{\lambda,0}\left(\sum_{n=0}^\infty p_n\;|n\rangle\langle n|\right)=\sum_{n=0}^\infty \left[T_\lambda(p)\right]_n\;|n\rangle\langle n|\;.
\end{equation}
\end{thm}
We recall that for any $E\ge0$, the thermal quantum Gaussian states $\hat{\omega}(E)$ corresponds to the geometric probability distribution $\omega(E)$ for the energy given by
\begin{equation}
{\omega(E)}_n = \frac{1}{E+1}\left(\frac{E}{E+1}\right)^n\;.
\end{equation}
We can then extend to the thinning all the results on the quantum-limited attenuator.

Let $p$ and $q$ be two probability distributions on $\mathbb{N}$.
We say that $p$ majorizes $q$, and write $p\succ q$, iff there exists a doubly stochastic infinite matrix $A$ such that \cite{marshall2010inequalities}
\begin{equation}
q_n = \sum_{k=0}^\infty A_{nk}\,p_k\qquad \forall\;n\in\mathbb{N}\;.
\end{equation}
The infinite matrix $A$ is doubly stochastic iff
\begin{equation}
A_{mn}\ge0\quad\forall\;m,\,n\in\mathbb{N}\;,\qquad \sum_{k=0}^\infty A_{nk} = \sum_{k=0}^\infty A_{kn} = 1\quad\forall\;n\in\mathbb{N}\;.
\end{equation}
The link with the majorization for quantum states of Definition \ref{def:maj} is the following.
\begin{thm}
The quantum state $\hat{\rho}$ majorizes the quantum state $\hat{\sigma}$ iff the probability distribution on $\mathbb{N}$ associated to the spectrum of $\hat{\rho}$ majorizes the probability distribution on $\mathbb{N}$ associated to the spectrum of $\hat{\sigma}$.
\end{thm}
Theorem \ref{thm:maj2} implies then
\begin{thm}\label{thm:majT}
For any $0\le\lambda\le1$ and any probability distribution $p$ on $\mathbb{N}$,
\begin{equation}
T_\lambda p \prec T_\lambda p^\downarrow\;,
\end{equation}
where $p^\downarrow$ is the decreasing rearrangement of $p$, i.e., $p_n^\downarrow = p_{\sigma(n)}$ for any $n\in\mathbb{N}$, where $\sigma:\mathbb{N}\to\mathbb{N}$ is a bijective function such that $p_{\sigma(0)} \ge p_{\sigma(1)} \ge \ldots \ge 0$.
\end{thm}

The Shannon entropy of the probability measure $p$ on $\mathbb{N}$ is the counterpart of the von Neumann entropy:
\begin{equation}
S(p) = -\sum_{n=0}^\infty p_n\ln p_n\;.
\end{equation}
The proof of Conjecture \ref{conj:MOE} for the one-mode quantum-limited attenuator \cite{de2016gaussian} implies
\begin{thm}\label{thm:EPnIT}
Geometric input probability distributions minimize the output Shannon entropy of the thinning among all the input probability distribution with a given Shannon entropy.
In other words, let $p$ be a generic probability distribution on $\mathbb{N}$, and let $\omega$ be the geometric probability distribution with the same Shannon entropy as $p$.
Then, for any $0\le\lambda\le1$
\begin{equation}
S(T_\lambda(p)) \ge S(T_\lambda(\omega)) = g\left(\lambda\,g^{-1}(S(p))\right)\;.
\end{equation}
\end{thm}

For any $p\ge1$, the $l^p$ norm of the sequence of complex numbers $\{x_n\}_{n\in\mathbb{N}}$ is
\begin{equation}
\left\|x\right\|_p = \left(\sum_{n\in\mathbb{N}}|x_n|^p\right)^\frac{1}{p}\;.
\end{equation}
For any $p,\,q\ge1$, the $p\to q$ norm of the thinning is
\begin{equation}
\|T_\lambda\|_{p\to q} = \sup_{0<\|x\|_p<\infty}\frac{\left\|T_\lambda x\right\|_q}{\|x\|_p}\;.
\end{equation}
The proof of Conjecture \ref{conj:pq} for the one-mode quantum-limited attenuator \cite{de2018pq} implies then
\begin{thm}\label{thm:pqT}
For any $p,\,q\ge1$, the $p\to q$ norm of the thinning is achieved by geometric probability distributions, i.e., for any $0\le\lambda\le1$,
\begin{equation}\label{eq:pqT}
\|T_\lambda\|_{p\to q} = \sup_{E\ge0}\frac{\left\|T_\lambda\omega(E)\right\|_q}{\left\|\omega(E)\right\|_p}\;.
\end{equation}
\end{thm}
\begin{rem}
The supremum in \eqref{eq:pqT} is
\begin{itemize}
\item finite and achieved for a finite $E\ge0$ if $1\le p<q$;
\item finite and asymptotically achieved in the limit $E\to\infty$ if $1<p=q$;
\item infinite and asymptotically achieved in the limit $E\to\infty$ if $1\le q<p$.
\end{itemize}
\end{rem}

\section{Quantum conditioning and the quantum Entropy Power Inequality}\label{sec:EPI}
\subsection{The quantum Entropy Power Inequality}
The first attempt to prove the Entropy Photon-number Inequality was through the quantum counterpart of the heat semigroup technique of the proof of the Entropy Power Inequality by Blachman and Stam.
However, this technique only leads to the quantum Entropy Power Inequality \cite{konig2014entropy,konig2016corrections,de2014generalization}, which has the same expression as the Entropy Power Inequality and provides a lower bound to the output entropy of the beam-splitter or of the squeezing in terms of the entropies of the two inputs.
Since this bound is strictly lower than the output entropy achieved by thermal Gaussian input states, the quantum Entropy Power Inequality is strictly weaker than the Entropy Photon-number Inequality.
\begin{thm}[quantum Entropy Power Inequality]\label{thm:qEPI}
For any $\lambda\ge0$ and any two $n$-mode quantum states $\hat{\rho}_A$ and $\hat{\rho}_B$ with a finite average energy,
\begin{equation}\label{eq:qEPI}
\exp\frac{S\left(\mathcal{B}_\lambda^{\otimes n}(\hat{\rho}_A\otimes\hat{\rho}_B)\right)}{n} \ge \lambda\exp\frac{S(\hat{\rho}_A)}{n} + \left|1-\lambda\right|\exp\frac{S(\hat{\rho}_B)}{n}\;.
\end{equation}
\end{thm}
\begin{rem}
The factors of $2$ in the exponents in the classical Entropy Power Inequality \eqref{eq:cEPI} do not appear in \eqref{eq:qEPI} because an $n$-mode quantum state is the counterpart of a random variable on $\mathbb{R}^{2n}$.
The coefficients in front of the exponentials in the right-hand side of \eqref{eq:qEPI} come from the coefficients in the transformation rules for the ladder operators \eqref{eq:defUlambda} and \eqref{eq:defUk}.
\end{rem}
The quantum Entropy Power Inequality was proposed by K\"onig and Smith \cite{konig2014entropy}, who proved it in the case $\lambda=\frac{1}{2}$ \cite{konig2014entropy,konig2016corrections}.
De Palma, Mari and Giovannetti extended the proof to any $\lambda\ge0$ \cite{de2014generalization}.
De Palma, Mari, Lloyd and Giovannetti proposed and proved an Entropy Power Inequality for the most general linear transformation of bosonic modes \cite{de2015multimode}.
Huber, K\"onig and Vershynina proposed and proved an Entropy Power Inequality for the quantum additive noise channels \cite{huber2017geometric}.

\subsection{Quantum conditioning}
In the classical scenario, the Shannon entropy of the random variable $A$ conditioned on the ``memory'' random variable $M$ with law $p$ is defined as the expectation value of the Shannon entropy of $A$ conditioned on the values assumed by $M$ \cite{cover2006elements}:
\begin{equation}\label{eq:defccS}
S(A|M) = \int S(A|M=m)\,\mathrm{d}p(m)\;.
\end{equation}
Let now $A$ and $M$ be quantum systems, and let us consider a quantum state $\hat{\rho}_{AM}$ on the joint system $AM$.
The definition \eqref{eq:defccS} cannot be brought to the quantum setting when $A$ is entangled with $M$, since conditioning on the values assumed by $M$ is not possible.
However, \eqref{eq:defccS} can be rewritten as
\begin{equation}\label{def:A|Mq}
S(A|M) = S(AM) - S(M)\;,
\end{equation}
that is the right definition for the quantum conditional entropy \cite{nielsen2010quantum,hayashi2016quantum,wilde2017quantum,holevo2013quantum} (see Ref. \onlinecite{tomamichel2015quantum} for a broad discussion).
We write $S(A|M)_{\hat{\rho}_{AM}}$ when the joint quantum state to which the conditional entropy refers is not clear from the context.
A striking feature of the quantum conditional entropy is that it can be negative, while the quantum entropy is always positive.

The correlation between two random variables or two quantum systems $A$ and $B$ are quantified by the (quantum) mutual information \cite{nielsen2010quantum,hayashi2016quantum,wilde2017quantum,holevo2013quantum}
\begin{equation}
I(A:B) = S(A) + S(B) - S(AB)\;.
\end{equation}
Both the classical and quantum versions of the mutual information are positive as a consequence of the subadditivity of the entropy \cite{nielsen2010quantum,hayashi2016quantum,wilde2017quantum,holevo2013quantum}.
The classical mutual information vanishes iff $A$ and $B$ are independent random variables.
Analogously, the quantum mutual information vanishes iff $\hat{\rho}_{AB} = \hat{\rho}_A\otimes\hat{\rho}_B$.

The conditional mutual information between $A$ and $B$ conditioned on the memory $M$ is
\begin{equation}
I(A:B|M) = S(A|M) + S(B|M) - S(AB|M)\;.
\end{equation}
The classical conditional mutual information is positive as a consequence of the expression \eqref{eq:defccS} for the conditional entropy and of the positivity of the mutual information \cite{cover2006elements}.
Also the quantum conditional mutual information is positive \cite{nielsen2010quantum,hayashi2016quantum,wilde2017quantum,holevo2013quantum}.
Since the quantum conditional entropy cannot be written as in \eqref{eq:defccS}, this result is highly nontrivial.
The classical conditional mutual information vanishes iff $A$ and $B$ are conditionally independent given the value of $M$.
The quantum conditional mutual information vanishes for all the joint quantum states of the following form \cite{hayden2004structure}
\begin{equation}\label{eq:ssasat}
\hat{\rho}_{ABM} = \bigoplus_{n=0}^\infty p_n\,\hat{\rho}_{AM_A^{(n)}}\otimes \hat{\rho}_{BM_B^{(n)}}\;,
\end{equation}
where $p$ is a probability distribution on $\mathbb{N}$ and each $\hat{\rho}_{AM_A^{(n)}}$ or $\hat{\rho}_{BM_B^{(n)}}$ is a quantum state on the Hilbert space $\mathcal{H}_A\otimes\mathcal{H}_{M_A^{(n)}}$ or $\mathcal{H}_B\otimes\mathcal{H}_{M_B^{(n)}}$, respectively, and where
\begin{equation}
\mathcal{H}_M = \bigoplus_{n=0}^\infty \mathcal{H}_{M_A^{(n)}}\otimes \mathcal{H}_{M_B^{(n)}}\;.
\end{equation}
If $A$, $B$ and $M$ have finite dimension, all the quantum states with vanishing conditional mutual information are of the form \eqref{eq:ssasat}.
The same property is believed to hold for infinite dimension, but this has not been proven yet.

A fundamental consequence of the positivity of the quantum conditional mutual information is the associated data-processing inequality, stating that discarding a subsystem always decreases the quantum conditional mutual information, i.e., for any quantum state on a joint quantum system $ABCM$
\begin{equation}\label{eq:dp}
I(AC:B|M) \le I(A:B|M)\;.
\end{equation}

\subsection{The quantum conditional Entropy Power Inequality}
Let $X$ and $Y$ be random variables with values in $\mathbb{R}^n$, and let $M$ be a random variable such that $X$ and $Y$ are conditionally independent given $M$.
Then, the expression \eqref{eq:defccS}, the Entropy Power Inequality \eqref{eq:cEPI} and Jensen's inequality imply the conditional Entropy Power Inequality \cite{de2018conditional}
\begin{equation}\label{eq:ccEPI}
\exp\frac{2S(X+Y|M)}{n} \ge \exp\frac{2S(X|M)}{n} + \exp\frac{2S(Y|M)}{n}\;.
\end{equation}
The inequality \eqref{eq:ccEPI} is saturated by any joint probability measure on $ABM$ such that, conditioning on any value $m$ of $M$, $A$ and $B$ are independent Gaussian random variables with proportional covariance matrices, and the proportionality constant does not depend on $m$.

Since the quantum conditional entropy cannot be expressed as in \eqref{eq:defccS}, the above proof does not go through in the quantum setting.
However, the following quantum conditional Entropy Power Inequality follows adapting the proof of the quantum Entropy Power Inequality.
\begin{thm}[quantum conditional Entropy Power Inequality]\label{thm:qcEPI}
Quantum Gaussian states minimize the output quantum conditional entropy of the beam-splitter and of the squeezing among all the input states where the two inputs are conditionally independent given the memory.
In other words, let $A$ and $B$ be $n$-mode Gaussian quantum systems, and let $M$ be a generic quantum system.
Let $\hat{\rho}_{ABM}$ be a joint quantum state with finite average energy on $AB$, finite $S(\hat{\rho}_M)$ and with $I(A:B|M)=0$, and let
\begin{equation}\label{eq:rhoCM}
\hat{\rho}_{CM} = \left(\mathcal{B}_\lambda^{\otimes n}\otimes\mathbb{I}_M\right)(\hat{\rho}_{ABM})\;,
\end{equation}
where $\lambda\ge0$ and $A$ and $B$ are the two inputs of the beam-splitter or of the squeezing.
Then,
\begin{equation}\label{eq:qcEPI}
\exp\frac{S(C|M)}{n} \ge \lambda\exp\frac{S(A|M)}{n} + \left|1-\lambda\right|\exp\frac{S(B|M)}{n}\;.
\end{equation}
Moreover, let $M$ be a $2n$-mode Gaussian quantum system of the form $M=M_AM_B$, where $M_A$ and $M_B$ are $n$-mode Gaussian quantum systems.
Then, for any $a,\,b\in\mathbb{R}$ there exists a sequence $\{\hat{\rho}_{ABM}^{(k)}\}_{k\in\mathbb{N}}$ of $4n$-mode quantum Gaussian states of the form $\hat{\rho}_{ABM}^{(k)} = \hat{\rho}_{AM_A}^{(k)} \otimes \hat{\rho}_{BM_B}^{(k)}$ such that
\begin{equation}
S(A|M)_{\hat{\rho}_{ABM}^{(k)}} = a\;,\qquad S(B|M)_{\hat{\rho}_{ABM}^{(k)}} = b
\end{equation}
for any $k\in\mathbb{N}$, and
\begin{equation}
\lim_{k\to\infty} \exp\frac{S(C|M)_{\hat{\rho}_{CM}^{(k)}}}{n} = \lambda\,\exp\frac{a}{n} + \left|1-\lambda\right|\exp\frac{b}{n}\;.
\end{equation}
\end{thm}
If $M$ is trivial, \eqref{eq:qcEPI} becomes the quantum Entropy Power Inequality.
The quantum conditional Entropy Power Inequality was first conjectured by K\"onig, who proved it in the case $0\le\lambda\le1$ for Gaussian input states \cite{koenig2015conditional}.
The general case was proven by De Palma and Trevisan \cite{de2018conditional}.
De Palma and Huber proved a conditional Entropy Power Inequality for the quantum additive noise channels \cite{huber2018conditional}.
The proofs of Refs. \onlinecite{de2018conditional,huber2018conditional} settled some regularity issues that affected the previous proofs of Refs. \onlinecite{konig2014entropy,de2014generalization,de2015multimode,koenig2015conditional}.

The proof of the quantum conditional Entropy Power inequality of Ref. \onlinecite{de2018conditional} is the quantum counterpart of the proof of the classical Entropy Power Inequality by Blachman and Stam based on the evolution with the heat semigroup.
Let $A$ be a $n$-mode Gaussian quantum system with ladder operators $\hat{a}_1,\,\ldots,\,\hat{a}_n$.
The displacement operator $\hat{D}(z)$ with $z\in\mathbb{C}^n$ is the unitary operator that displaces the ladder operators:
\begin{equation}
{\hat{D}(z)}^\dag\,\hat{a}_i\,\hat{D}(z) = \hat{a}_i + z_i\,\hat{\mathbb{I}}\qquad i=1,\,\ldots,\,n\;.
\end{equation}
The quantum heat semigroup is the quantum Gaussian channel generated by a convex combination of displacement operators with a Gaussian probability measure:
\begin{equation}\label{eq:heat}
\mathcal{N}_t(\hat{\rho}) = \int_{\mathbb{C}^n}\hat{D}\left(\sqrt{t}\,z\right)\,\hat{\rho}\,{\hat{D}\left(\sqrt{t}\,z\right)}^\dag\,\mathrm{e}^{-|z|^2}\,\frac{\mathrm{d}z}{\pi^n}\;,\quad \mathcal{N}_0=\mathbb{I}\;,\quad \mathcal{N}_t\circ\mathcal{N}_{t'}=\mathcal{N}_{t+t'}\quad \forall\;t,\,t'\ge0\;.
\end{equation}
This is the quantum counterpart of the classical heat semigroup acting on a probability density function $f$ on $\mathbb{C}^n$:
\begin{equation}
(\mathcal{N}_t f)(w) = \int_{\mathbb{C}^n} f\left(w-\sqrt{t}\,z\right)\,\mathrm{e}^{-|z|^2}\,\frac{\mathrm{d}z}{\pi^n}\;,\qquad w\in\mathbb{C}^n
\end{equation}
Let $\hat{\rho}_{AM}$ be a joint quantum state on $AM$.
The quantum conditional Fisher information of the state $\hat{\rho}_{AM}$ is the rate of increase of the quantum conditional mutual information between $A$ and $Z$ when the system $A$ is displaced by $\sqrt{t}\,Z$ according to \eqref{eq:heat}.
In other words, for any $t>0$, let $\hat{\sigma}_{AMZ}(t)$ be the probability measure on $\mathbb{C}^n$ with values in quantum states on $AM$ such that
\begin{equation}
\mathrm{d}\hat{\sigma}_{AMZ}(z,t) = \hat{D}\left(\sqrt{t}\,z\right)\;\hat{\rho}\;{\hat{D}\left(\sqrt{t}\,z\right)}^\dag\mathrm{e}^{-|z|^2}\frac{\mathrm{d}z}{\pi^n},\quad \int_{\mathbb{C}^n}\mathrm{d}\hat{\sigma}_{AMZ}(z,t) = (\mathcal{N}_t\otimes\mathbb{I}_M)(\hat{\rho}_{AM})\;.
\end{equation}
Then, the quantum conditional Fisher information of $\hat{\rho}_{AM}$ is
\begin{equation}
J(A|M)_{\hat{\rho}_{AM}} = \left.\frac{\mathrm{d}}{\mathrm{d}t}I(A:Z|M)_{\hat{\sigma}_{AMZ}(t)}\right|_{t=0}\;.
\end{equation}
The quantum de Bruijn identity links the quantum conditional Fisher information to the time derivative of the conditional entropy along the heat semigroup.
\begin{lem}[quantum de Bruijn identity]
\begin{equation}
J(A|M)_{\hat{\rho}_{AM}} = \left.\frac{\mathrm{d}}{\mathrm{d}t}S(A|M)_{(\mathcal{N}_t\otimes\mathbb{I}_M)(\hat{\rho}_{AM})}\right|_{t=0}\;.
\end{equation}
\end{lem}
The first part of the proof of the quantum conditional Entropy Power Inequality is proving the following quantum conditional Stam inequality, which provides an upper bound to the quantum conditional Fisher information of the output of the beam-splitter or of the squeezing in terms of the quantum conditional Fisher information of the two inputs.
\begin{thm}[quantum conditional Stam inequality]
Let $\hat{\rho}_{ABM}$ be a quantum state on $ABM$ with finite average energy, finite $S(M)$ and $I(A:B|M)=0$, and let $\hat{\rho}_{CM}$ be as in \eqref{eq:rhoCM}.
Then, for any $\lambda\ge0$ the quantum conditional Stam inequality holds:
\begin{equation}\label{eq:Stam}
\frac{1}{J(C|M)_{\hat{\rho}_{CM}}} \ge \frac{\lambda}{J(A|M)_{\hat{\rho}_{AM}}} + \frac{|1-\lambda|}{J(B|M)_{\hat{\rho}_{BM}}}\;.
\end{equation}
\end{thm}
The quantum conditional Stam inequality follows from the data processing inequality for the quantum conditional mutual information \eqref{eq:dp}, and implies that the quantum conditional Entropy Power Inequality does not improve along the evolution with the heat semigroup.
Then, the proof of the Entropy Power Inequality is concluded if we show that it becomes asymptotically an equality in the infinite time limit.
This is achieved by proving that the quantum conditional entropy has an universal scaling independent on the initial state in the infinite time limit under the evolution with the heat semigroup.
\begin{lem}
For any joint quantum state $\hat{\rho}_{AM}$ with finite average energy and finite $S(M)$,
\begin{equation}
S(A|M)_{(\mathcal{N}_t\otimes\mathbb{I}_M)(\hat{\rho}_{AM})} = n\ln t + n + o(1)\qquad \text{for}\;t\to\infty\;.
\end{equation}
\end{lem}
The proof of this scaling is based on the following more general result.
\begin{thm}\label{thm:mince}
Let $A$, $B$ be quantum Gaussian systems with $m$ and $n$ modes, respectively, and let $\Phi:A\to B$ a quantum Gaussian channel.
Then, for any quantum system $M$ and any quantum state $\hat{\rho}_{AM}$ on $AM$ with finite average energy and finite $S(M)$,
\begin{equation}
S(B|M)_{(\Phi\otimes\mathbb{I}_M)(\hat{\rho}_{AM})} \ge \lim_{E\to\infty}S(B|A')_{(\Phi\otimes\mathbb{I}_{A'})(\hat{\tau}_{AA'}(E))}\;,
\end{equation}
where $A'$ is a Gaussian quantum system with $m$ modes, and for any $E\ge0$, $\hat{\tau}_{AA'}(E)$ is a pure state such that its marginal on $A$ is the thermal Gaussian state ${\hat{\omega}(E)}^{\otimes m}$.
\end{thm}
We mention that a result similar to Theorem \ref{thm:mince} has been proven in the scenario with a constraint on the average energy of the system $A$ \cite{de2018minimum}.

\section{Conclusions and perspectives}\label{sec:concl}
The optimization problems of functional analysis whose solutions are Gaussian functions have stimulated to conjecture that quantum Gaussian states are the solution to the quantum counterparts of these optimization problems.
These conjectures play a key role in quantum information theory, since they are necessary to prove the converse theorems for many communication scenarios with quantum Gaussian channels.
We have reviewed the state of the art in the proof of these conjectures.
In the case of one-mode quantum Gaussian channels, they are almost all solved, with the exceptions of the Entropy Photon-number Inequality and the sharp Young's inequality for the beam-splitter.
On the contrary, there are only very few results for multi-mode quantum Gaussian channels.
In this scenario, both the constrained minimum output entropy conjecture (Conjecture \ref{conj:MOE}) and the multiplicativity of the $p\to q$ norms with $1<p<q$ (Conjecture \ref{conj:pq}) are still completely open challenging problems, and we hope that this review will set the ground for their solution.

Quantum Gaussian channels also constitute a bridge between continuous and discrete classical probability.
Indeed, on the one hand their properties are very similar to the properties of Gaussian integral kernels, with quantum Gaussian states playing the role of Gaussian probability measures.
On the other hand, the quantum states diagonal in the Fock basis of a one-mode Gaussian quantum system are in a one-to-one correspondence with the probability measures on $\mathbb{N}$.
This correspondence establishes a bridge between Gaussian quantum system and discrete probability.
The role of quantum Gaussian states is here played by the geometric probability distributions.
These distributions turn out to be the solution to many optimization problems involving the thinning, which is the discrete analogue of the rescaling of a real random variable.

We then hope that this review will stimulate even more cross-fertilization among functional analysis, discrete probability and quantum information.

\section*{Acknowledgements}
We thank Elliott Lieb and Rupert Frank for a careful reading of the review and useful comments.

GdP acknowledges financial support from the European Research Council (ERC Grant Agreements Nos. 337603 and 321029), the Danish Council for Independent Research (Sapere Aude) and VILLUM FONDEN via the QMATH Centre of Excellence (Grant No. 10059), and the Marie Sk\l odowska-Curie Action GENIUS (Grant No. 792557).

\includegraphics[width=0.05\textwidth]{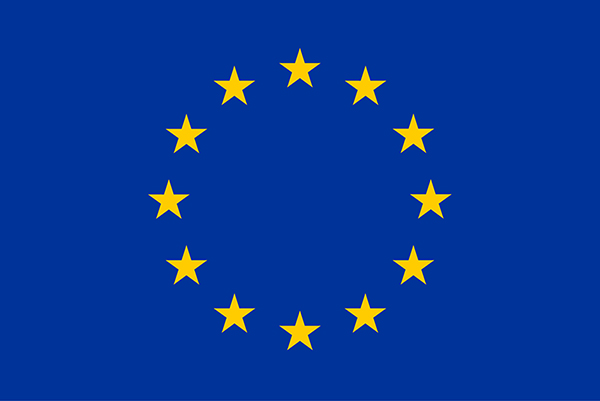}
This project has received funding from the European Union's Horizon 2020 research and innovation programme under the Marie Sk\l odowska-Curie grant agreement No. 792557.

\section*{Author Contributions}
The authors equally contributed to conceive the review and prepare the sketch.
The main text was written by GdP.
All the authors helped to shape and review the manuscript.

\bibliographystyle{plain}
\bibliography{biblio}
\end{document}